\documentclass[traditabstract]{aa}

\usepackage{natbib}
\usepackage{graphics}
\usepackage{graphicx}
\usepackage{epsfig}
\usepackage{amssymb}

\usepackage{txfonts}

\usepackage{epsf}
\usepackage[T1]{fontenc}
\usepackage{color}
\definecolor{red}{rgb}{0.7,0,0}
\definecolor{blue}{rgb}{0,0,0.7}
\bibpunct{(}{)}{;}{a}{}{,}

\def\kir{$\chi^2_{\rm red}$}

\def\gx{GX~339$-$4}

\newcommand{\rxte}{\textsl{RXTE}}
\newcommand{\integral}{\textsl{INTEGRAL}}
\newcommand{\swift}{\textsl{Swift}}

\begin{document}
\title{Overview of an Extensive Multi-wavelength Study of GX~339$-$4 during the 2010 Outburst}

\author{M. Cadolle Bel\inst{1}, J. Rodriguez\inst{2}, P. D'Avanzo\inst{3}, D. M. Russell\inst{4}, J. Tomsick\inst{5}, 
S. Corbel\inst{2}, F. Lewis\inst{6}, F. Rahoui\inst{7}, M. Buxton\inst{8}, P. Goldoni\inst{9} and E. Kuulkers\inst{1}}

\offprints{Dr. Cadolle Bel: Marion.Cadolle@sciops.esa.int}

\institute{ESAC, ISOC, Villa{\~n}ueva de la Ca{\~n}ada, Madrid, Spain
\and Laboratoire AIM, UMR 7158, CEA/DSM - CNRS - Universit\'e Paris Diderot, IRFU/SAp, Gif-sur-Yvette, France
\and INAF, Osservatorio Astronomico di Brera, Merate, Italy 
\and Univ. of Amsterdam (Anton Pannekoek Institute), The Netherlands
\and SSL/Univ. California, Berkeley, USA
\and Faulkes Telescope Project, Univ. of Glamorgan, Wales
\and Astronomy Department, Harvard Univ. \& Harvard-Smithsonian Center for Astrophysics, Cambridge, USA
\and Astronomy Department, Yale Univ., USA
\and Laboratoire APC, UMR 7164, CEA/DSM - CNRS - Universit\'e Paris Diderot, IRFU/SAp, Paris, France}

\date{Accepted on 2011, September 20}
\authorrunning{M. Cadolle Bel et al.}
\titlerunning{Multi-wavelength studies of \gx}

\abstract
{The microquasar \gx\ experienced a new outburst in 2010: it was observed simultaneously at various wavelengths from radio up to soft $\gamma$-rays. We focused on observations that are quasi-simultaneous with those made with the \integral\ and \rxte\ satellites: these were collected in 
2010 March--April during our \integral\ Target of Opportunity 
program, and during some of the 
other \integral\ observing programs with \gx\ in the field-of-view.
X-ray transients are extreme systems that often harbour a black hole, 
and are known to emit throughout the whole electromagnetic spectrum when in outburst. The 
goals of our program are to understand the evolution of the physical processes close to the 
black hole and to study the connections between the accretion and ejection.
We analysed radio, NIR, optical, UV, X-ray and soft $\gamma$-ray observations. We studied 
the source evolution in detail by producing light curves, hardness-intensity diagrams and spectra. We 
fitted the broadband data with phenomenological, then physical, models to study the emission coming 
from the distinct components.
Based on the energy spectra, the source evolved from the canonical hard state to the canonical
soft state. The source showed X-ray spectral variations that were correlated with changes in radio, NIR and optical emission. 
The bolometric flux increased from 0.8 to 2.9 
$\times 10^{-8}$ erg cm$^{-2}$ s$^{-1}$ while the relative flux and contribution of the hot medium 
globally decreased. Reprocessing in the disc was likely to be strong at the end of our observations.
The source showed a behaviour similar to that of previous outbursts, with some small deviations 
in the hard X-rays parameters' evolution. The radio, NIR and optical emission from jets was detected, and seen to 
fade as the source softened. The results are discussed within the context of disc and jet models.}


\maketitle

\section{Introduction}
\label{intro}

\begin{figure*}[t!]
\centering\includegraphics[angle=270,width=0.9\linewidth]{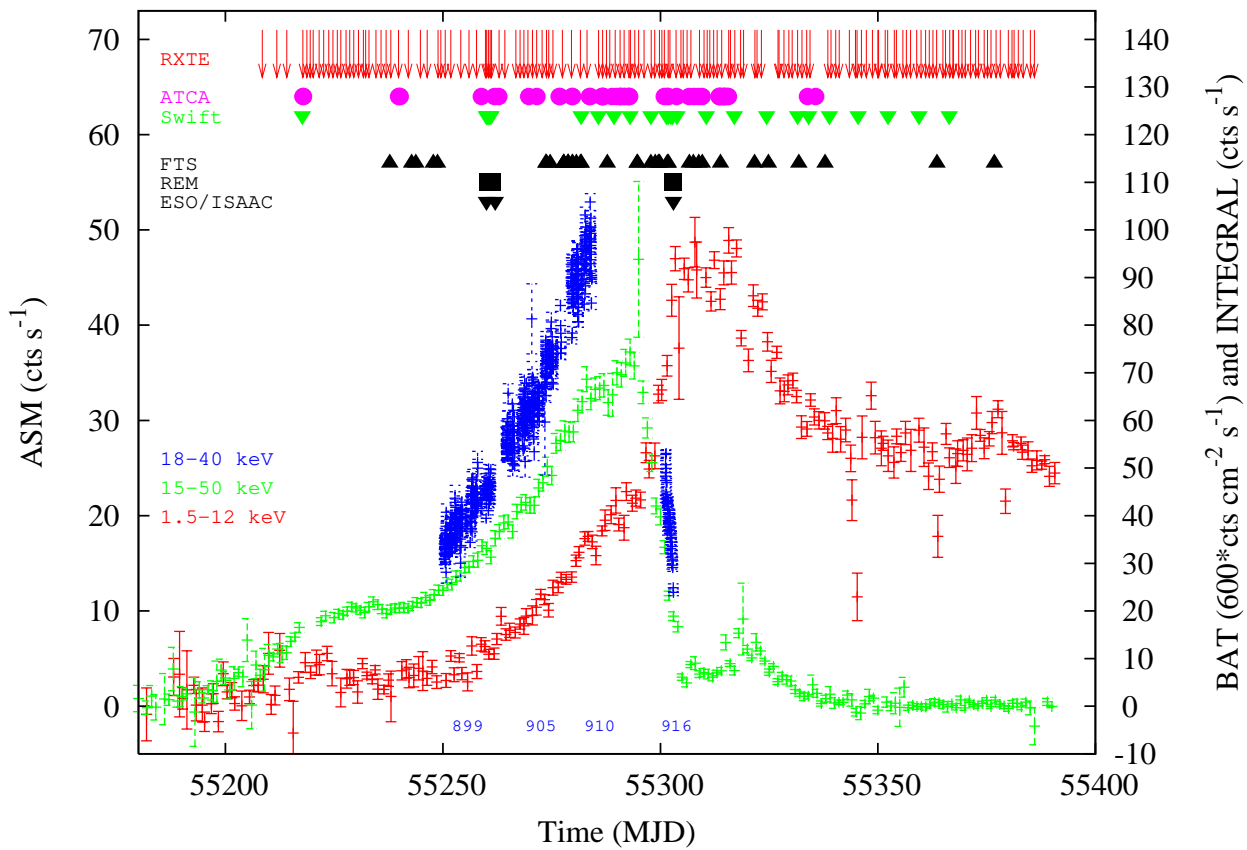}
\caption{\label{liste_obs}{Daily {{\emph {RXTE}/}}ASM (red), {\emph{Swift}}/BAT (green) 
and {\emph{INTEGRAL}}/ISGRI (per pointing, blue) light curves of \gx\ during the 2010 outburst. 
Other observations performed with ATCA, the FTS, the REM/ROSS, REMIR and 
the ESO/ISAAC telescopes are indicated.The blue numbers near the bottom of the Figure are 
{\emph{INTEGRAL}} revolution numbers (see Table~\ref{tab:log-gen} for details).}}
\end{figure*}

X-ray transients (XTs) are accreting low-mass
X-ray binaries (LMXBs) that spend most of their time in a faint, quiescent state. 
They undergo large amplitude outbursts with rise times of only a few days or weeks (or months in the case of \gx), 
with typical recurrence periods of many years \citep{Tanaka:1996}. 
It is commonly accepted that the outburst of an XT is the result of 
an accretion disc instability \citep{Frank2002}. This process triggers a transition from a quiescent 
state with a very low mass accretion rate ($\dot{m}$ $<<$ 0.01$\dot{M}$$_{Edd}$), where the source 
 activity is very faint or undetected (luminosity $\leq$ 10$^{33}$ erg s$^{-1}$), 
 to active states. 
 During outbursts, the optically thick and geometrically thin accretion disc has a varying inner radius \citep[e.g.,][]{tomsick09}
and temperature, emitting at typical X-ray energies of $\sim$1~keV. 
The inner regions of the disc (and the compact object) are embedded
in a hot and tenuous medium often referred to as the corona, where soft X-ray photons 
originating in the disc undergo inverse Comptonization. This results in a power law 
spectrum that has been detected up to a few MeV in some sources \citep{Grove98, zdz04,cad06,laurent07}. The exact 
origin of this high-energy power law spectrum is, however, subject to debate, and part of the high-energy emission 
could instead arise from synchrotron emission from a relativistic jet, that is usually seen in the radio domain \citep{Markoff:2005}. 
Note that \citet{laurent2011}, through the discovery of polarized $\gamma$-ray emission, recently showed that both
inverse Comptonisation and synchotron emission explained the 20~keV to $\sim2$~MeV spectrum of the persistent X-ray 
binary Cygnus X-1.\\ 
The spectral characteristics are coupled with different levels of variability, quasi 
periodic oscillations (QPOs) observed in the power density spectrum (PDS) (e.g., \citealt{Belloni:2001, Belloni:2005}) 
and with the properties of the jet emission at radio, optical and IR frequencies \citep[see, e.g.,][]{Russell:2006, coriat09}. 
Based on the relative strengths of each spectral component, the degree of variability 
and shape of the PDS, and properties of the radio emission, 
different spectral states have been defined \citep[e.g.,][]{McClintock:2006,Homan:2005}. 
The two main canonical spectral states are the low/hard state (LHS) and the high/soft state (HSS).
 In the LHS, the X-ray energy spectrum is dominated by hard power law-like emission and
the PDS shows a high level of variability dominated by 
a strong ($\sim$30$\%$ fractional rms) band-limited noise, associated with the presence of low-frequency 
QPOs. The LHS exhibits a continuously launched compact jet, 
whose radio-spectrum is flat \citep{Corbel:2000, Corbel:2003, Gallo:2003, Gallo:2006, Fender:2004}.
In the HSS, thermal emission from the disc dominates the spectrum, and the PDS shows only a weak power law noise component. 
No core radio emission is detected, and radio detections made during the HSS are likely to originate in 
discrete ejections launched previously \citep[see, e.g., ][]{fender09}. 
Other states have been identified and both are intermediate flavors of the canonical ones. These are 
the HIMS (hard intermediate state) and SIMS (soft intermediate state) as defined in \citet{Bell10}.

\noindent \gx\ is a recurrent XT with regular outbursts, shown to be a $\sim$7 solar mass BH from dynamical 
studies \citep{hynes2003,munoz2008} with well-sampled X-ray (e.g., \citealt{Belloni:2005}) and 
multi-wavelength coverage during outbursts (e.g., \citealt{Corbel:2000,Homan:2005b,coriat09}). 
On January 3, 2010, \gx\ entered a new outburst \citep{yam}. We triggered {\emph{INTEGRAL}} during the 
initial hard X-ray phase \citep{tomsick2010, Prat10}; subsequently, we 
triggered the second part of our {\emph{INTEGRAL} campaign on \gx\ (see Fig.~\ref{liste_obs}) in its 
declining hard X-ray phase \citep{cad10}. The source increased 
in intensity in soft X-rays while it kept declining in hard X-rays (as announced in \citealt{shap10,yu10}), 
exiting the LHS \citep{Bux10} towards a softer state.\\
Here, we report the results of our \integral\ observations 
of \gx\ together with \swift, \rxte\ and 
UV/optical/NIR/radio data. We start with a description of the available data and of the analysis procedures employed in Sect.~\ref{observations}.
Results are presented in Sect.~\ref{results}, followed by their interpretations and modelling 
in Sect.~\ref{discussion}. We summarize our studies in Sect.~\ref{summary}.

\begin{table*}[htbp]
\begin{center}
\caption{\label{tab:log-gen} Log of the \gx\ observations analysed in this paper.}
\begin{tabular}[h]{c c r@{--}l r@{--}l c c}
\hline \hline
Observatory &Instrument & \multicolumn{2}{c}{Bandpass} & \multicolumn{2}{c}{Period} & Observation Type \\
& & \multicolumn{2}{c}{}  & \multicolumn{2}{c}{(MJD$-$55000)} &  &\\
\hline

\integral$^a$     & IBIS/ISGRI   & 18    & 200~keV & 237.5 & 303.3                &  ToO (5$\times$5$^b$, Hex$^c$)\\
 & JEM-X & 5 & 25~keV & 237.5 & 303.3 &  ToO \\ 
\rxte$^a$         & PCA          & 3     & 25~keV  & 208.5 & 345.4          & Public\\
\swift$^d$    & XRT          & 0.3   & 10~keV   & 217.7 & 352.3       & Public \\
    & BAT          & 20   & 100~keV   & 217.7 & 352.3        &  Public \\
        & UVOT          & 180   & 600 nm   &  217.7 & 352.3        & Public \\
REM$^d$       & ROSS, REMIR         & 550   & 2159~nm  &  260.2       & 261.4 and 303   & ToO \\
FTS$^d$ & Merope & 460 & 883~nm & 237.7 & 337.8 &  Public \\
ATCA$^d$ & - & 5.5 & 9 GHz & 261 & 263 & ToO \\
\hline
\end{tabular}
\end{center}
\begin{list}{}{}
\item[$^a$] Interrupted observations: almost all Rev. ($\sim$ 3 days) between \# 895--916 for \integral; Galactic Bulge monitoring for \rxte.
\item[$^b$]5$\times$5 dither pattern around the nominal target location.
\item[$^c$] Hexagonal pattern around the nominal target location.
\item[$^d$]Snap-shot observations with the specified instruments or receivers.
\end{list}
\end{table*}
\section{Observations and data reduction}
\label{observations}

Table~\ref{tab:log-gen} summarizes the observations analysed in this paper, providing instrument details, energy ranges 
dates and modes. 
Figure~\ref{liste_obs} shows the \rxte/ASM, {\it SWIFT}/BAT X-ray and {\emph{INTEGRAL}} light curves of the outburst, 
with the time of the other multi-wavelength observations indicated.

\subsection{\integral}

Target of Opportunity observations were performed during 55259--55261 and 55301--55303 MJD, corresponding 
to revolutions \# 902 and \# 916 respectively (hereafter, Rev.: an \integral\ orbit around the Earth lasts
$\sim$3~days). We also used data from \integral\ programs of observations of the 
Galactic Bulge \citep{Kuulkers:2007}, RX 
J1713.7-3946 and the Galactic disc (performed between 
55237--55285 MJD, Rev. \# 895--910).\\
The IBIS/ISGRI and JEM-X data were reduced using standard analysis
procedures of the Off-Line Scientific Analysis {\tt OSA~9.0} following standard procedures 
(e.g., \citealt{Rodriguez:2008b,cad09}) 
for the production of images, spectra and light curves}. Systematic errors of 2\% were added for both
JEM-X (in the 5--25~keV range) and ISGRI (in the 18--400~keV range). We used
the maps, the response matrices and the off-axis and background corrections
from {\tt OSA~9.0}. We checked that the spectral index did not change by
more than 2\% during a single revolution by carrying out spectral analysis for each 
pointing. The low level of spectral variation over a single revolution allowed us to average all spectra belonging 
to the same revolution, which permitted the signal 
to noise ratio to be improved. The ISGRI light curves are shown in Figures~\ref{liste_obs} and~\ref{lcall} (fourth panel down). 
We could only perform JEM-X analysis on our ToO data as the source was outside the
field-of-view (FOV) of JEM-X in the other programs. Since the \rxte/PCA
instrument observed the source more frequently and had a higher sensitivity
than JEM-X, the JEM-X data were not included in the broad-band spectra so as
to be consistent over all our data sets. However, we verified that the best-fit
spectral parameters using JEM-X and PCA were consistent within the error
bars.

\subsection{\rxte}

We analysed all available observations, taken about once every two days from 
$\sim$ 55217 to 55311 MJD. Each observation lasted between $\sim$1 and 5~ks. 
The \rxte\ data were reduced with the {\tt HEASOFT} software package
v6.10, following standard procedures \citep[see][]{Rodriguez:2008b}. 
Energy spectra were only extracted from the top layer of PCA detector 2.
We used the latest calibration files provided by the \rxte\ guest observer
facility. A level of 0.6\% systematic error was adopted for \gx\ to
account for uncertainties in the PCA response 
\citep{Jahoda:2006}. The resultant \rxte/PCA spectra of a single observation were fitted simultaneously 
between 3--25~keV. Note that we also added the IBIS/ISGRI data when
available; one \integral\ revolution corresponds typically to three distinct
\rxte\ observations (we checked the spectral parameters were not changing significantly and could be 
averaged). ASM and PCA light curves are shown in Figure~\ref{lcall} (first and second panels down).

\subsection{\swift}

All the Swift/XRT observations discussed in this paper were made in 
Window Timing (WT) mode to avoid pile up due to the source brightness. 
Level 2 event files were produced following standard processing and screening procedures 
with {\tt{xrtpipeline}} (v0.12.6). 
The source and background spectra were then extracted from these files within {\tt{xselect}} from the 
grade 0 events only. Source spectra were obtained within a 40-pixel diameter circle centred on the 
source position, while background spectra were obtained using a similar sized region at an off-axis position.
An ancillary response file taking into account the exposure map of the observation was generated for 
each spectrum with {\tt{xrtmkarf}}, and the latest version of the redistribution matrix file was used in the fitting process. The original spectra have been rebinned by a factor of 4 before the fitting process. Besides, the BAT light curve 
(publicly available on-line) is shown in Figures~\ref{liste_obs} and~\ref{lcall} (third panel down).\\
In addition, data were collected from the \swift/UVOT 
(UV and optical) instrument, with MJD spanning from 55217 to 55310 simultaneously with 
{\emph{INTEGRAL}}. The six UVOT filters span the wavelength range 
1800--6000 $\AA$. The camera has a 17' $\times$ 17' FOV, resulting in images of 2048 $\times$ 2048 pixels \citep{romiet05}. 
UVOT images are combined using the standard UVOT {\tt uvotimsum} routine and magnitudes calculated using the 
{\tt uvotsource} command as provided by HEASARC using an aperture of 6'' \citep[see][]{poolet08}. 
Detections (at a significance of 3$\sigma$) are discussed in Section~\ref{discussion}.

\subsection{REM and FTS}

Optical and NIR observations were performed with the REM (rapid eye mounting) telescope 
\citep[see, e.g., ][]{Zerb,chin,Cov} equipped with the
ROSS optical spectrograph/imager and the REMIR NIR camera.
Observations of GX 339-4 were carried out around 55260.3, 
55261.3 and 55303.3 MJD. 
Image reduction was carried out by following standard procedures: 
subtraction of an averaged bias frame and division by a normalized flat frame. 
Astrometry was performed using the USNOB1.0\footnote{http://www.nofs.navy.mil/data/fchpix/} and the 
2MASS\footnote{http://www.ipac.caltech.edu/2mass/} catalogues.
Aperture photometry was performed with the SExtractor package \citep[]{Bertin96}
for all the objects in the field to analyse the REM data. 
The calibration was done against Landolt standard stars for the optical filters and against the 2MASS catalog for NIR 
filters. In order to minimize any systematic effect, we performed differential 
photometry with respect to a selection of local isolated and non-saturated 
standard stars.\\
Optical data have also been collected using the 2-metre robotic Faulkes Telescope South (FTS) 
located at Siding Spring, New South Wales, Australia using the Merope camera (EM03) coupled 
with an E2V CCD42-40DD CCD. This produces a 4.'7 $\times$ 4.'7 FOV and 
images of 2048 $\times$ 2048 pixels binned 2 $\times$ 2 to give 1024 $\times$ 1024 pixels at 
0.278 arcsec pixel$^{-1}$. Science images were produced using an automatic pipeline which 
de-biases and flat-fields the raw images\footnote{http://telescope.livjm.ac.uk/Info/TelInst/Inst/RATCam/\#pipe}. 
Generally, 200 second exposures were acquired in each of the Bessel-$V$, Bessel-$R$ and SDSS-$i'$ filters,
approximately weekly as part of a long-term monitoring campaign on \gx\ and 34 other LMXBs \citep{lewiet08}. 
On some dates, multiple images were taken (mostly in the $i'$ band) to investigate variability on minute-timescales. 
Seeing values range from 0.''8 to 3.''2. Images were discarded if the signal-to-noise ratio was low 
(often due to thin cloud), or if the tracking or focus were poor. In total, there are 338 usable images taken 
within the time range of interest. 
We performed aperture photometry of \gx\ and four nearby field stars using APPHOT in \small IRAF\normalsize. 
In addition, we used a non-variable star, which was fainter or of similar magnitude to \gx\ to assess its 
level of variability. A fixed aperture radius of 6 pixels along with a background annulus of 10--20 pixels 
was adopted for all stars in all filters. Flux calibration in $V$, $R$ and $i'$ bands was achieved using photometry 
of the standard star field RU 149, from the list of Landolt photometric stars \citep{land92}, which are observed regularly 
by FTS. We used the stars A, B, C, D and F as catalogued in the star field RU 149. The $i'$ magnitudes of the stars in 
the RU 149 field were estimated from the known $R$ and $I$ band magnitudes using the conversion of \cite{jordet06}. 
Uncertainties were calculated from the range of measurements from the RU 149 field stars (5 stars $\times$ 2 per filter 
= 10 measurements per filter). We used the small differences between our measurements and these reported 
measurements to apply a small correction to the $V$, $R$ and $i'$ calibration of the \gx\ 
field stars to achieve a more accurate calibration. FTS  light curves are shown in Figure~\ref{lcall} (lower three panels).\\

\subsection{Radio}

\gx\ was frequently observed at radio frequencies with the Australia Telescope Compact Array (ATCA) and the new CABB back-end. ATCA observed the source several times (PI Corbel) during the outburst. We concentrated on the simultaneous observations gathered with \integral~and ground-based instruments at 55261.89 and 55262.91 MJD. Fluxes were, at 5.5 and 9.0 GHz, respectively, 9.02$\pm$0.10 mJy, 9.56$\pm$0.05 mJy then 8.16$\pm$0.05 mJy and 7.94$\pm$0.10 mJy.

\section{Results}
\label{results}

The following sub-sections refer to the X/$\gamma$-ray, UV, optical and NIR light curves 
shown in Figs.~\ref{lcall} as well as to the HID (Fig.~\ref{hid}) and radio data. 
Some spectra are shown in Figs.~\ref{spe1} and ~\ref{spe2}.

\subsection{Light curves}

\begin{figure}
\centering
\includegraphics[width=9.2cm]{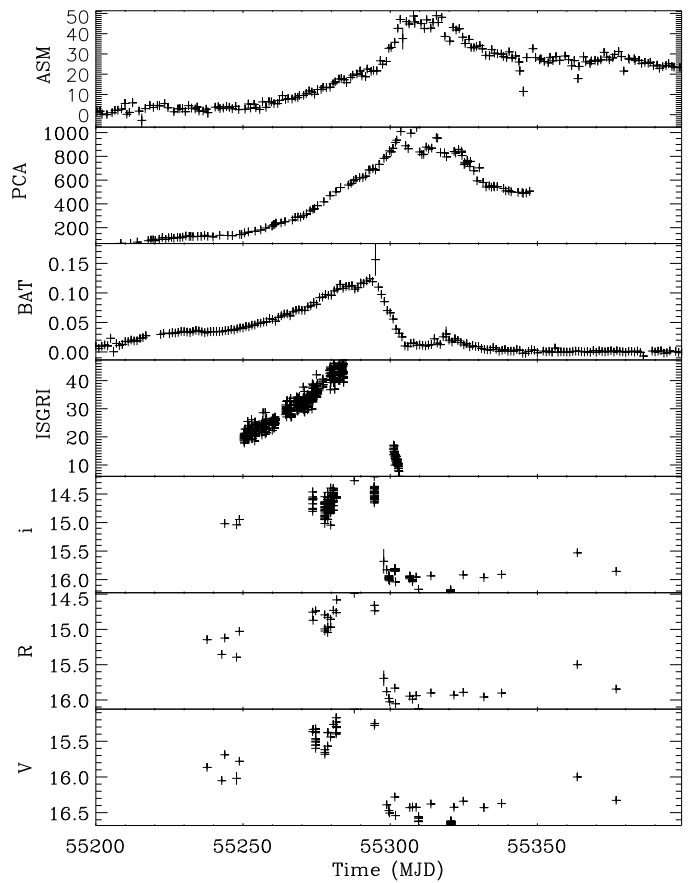}
\caption{\label{lcall}~{\emph{RXTE}}/ASM (cts s$^{-1}$ in 1.5--12 keV), 
PCA (cts s$^{-1}$ in 3--30 keV), {\emph{Swift}}/BAT (cts cm$^{-2}$ 
s$^{-1}$ in 15--50 keV), {\emph{INTEGRAL}}/ISGRI (cts s$^{-1}$ in 40--80 keV) 
and FTS light curves (filters $V$, $R$ and $i'$) obtained from 55200 to 55400 MJD.} 
\end{figure}

Fig.~\ref{lcall} presents the light curves obtained with the ASM (2--10 keV), PCA (3--30), BAT (15--40 keV), ISGRI (40--80 keV) and FTS ($V$, $R$ and $i'$ magnitudes) instruments. The source shows the typical behaviour of a XT: the outburst started in hard X-rays probably around $\sim$55199 MJD. The light curve morphologies at the peak of BAT and ASM are quite different. The hard X-rays dropped before the soft X-ray peak: around 55295 MJD, the hard X-rays reached their peak and the emission started to decrease (with a secondary peak at 55319 MJD) while the source continued to increase in soft X-rays until 55327 MJD (i.e., more than a month later), before decreasing. This behaviour is indicative of a state transition from a hard to a softer state. The optical (e.g., Fig.~\ref{lcall}) also dropped at the start of the transition, as reported in previous outbursts of \gx\ \citep{Homan:2005b,coriat09}.\\
While during 55259.9--55261.1 MJD (Rev. \# 902) the source was still in a hard state, during 55301.1--55303.3 MJD (Rev. \# 916) 
\gx\ was extremely bright in the Swift/XRT ($>$ 400 cts s$^{-1}$) instrument. The JEM-X fluxes were 
around 370 mCrab and 240 mCrab in, respectively, the 3--10 and 10--20 keV energy bands, 
while the ISGRI fluxes were varying around 190 mCrab and 130 mCrab in the 20--40 keV and 40--80 keV 
energy bands. These numbers show the X-ray source was softening. 

\subsection{Hardness intensity diagram}
\label{secthid}

\begin{figure}
\centering
\includegraphics[angle=270,width=8.5cm]{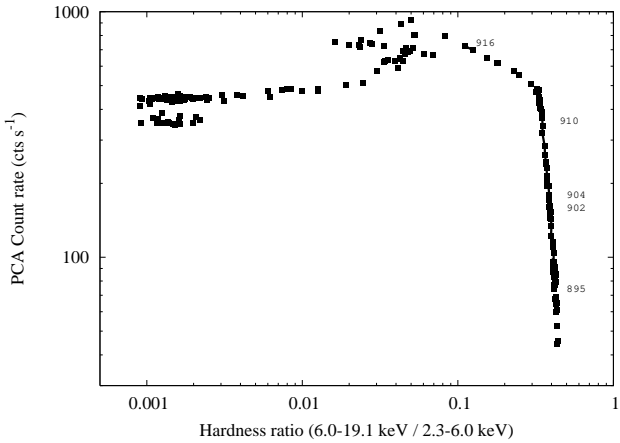}
\caption{\label{hid}~{HID of \gx\ (logarithmic scale) obtained with the top layer of \emph{RXTE}}/PCA (2.3--19.1 keV, detector 2) between 55208 and 55347 MJDs (the numbers indicate some of the INTEGRAL revolutions). The source 
globally evolves from bottom to top, and from right to left. }
\end{figure}

To study the X-ray spectral and flux evolution during the outburst, we produced a
hardness-intensity diagram (HID) with \emph{RXTE}/PCA (Fig.~\ref{hid}) similar to 
those widely used in the literature \citep[e.g.,][]{Fender:2004}. There was no {\it INTEGRAL} coverage 
from the start of the outburst to the peak, hence we focused on the main part of the 
outburst simultaneously with our \integral\ data (the HID extends slightly beyond the main data log analysed 
in this paper, to illustrate the evolution of the source after our campaign). 
\gx\ followed the usual path of XTs in
outburst on this diagram (see, e.g., \citealt{Belloni:2005}), with some small deviations in the usual track.

\subsection{X-ray and $\gamma$-ray spectra}
\label{xgamspec}

\begin{table*}[t!]
\begin{center}
\caption{\label{tab:para}{Best-fit spectral parameters around the period of our main {\emph {INTEGRAL}} observations. }}
\begin{tabular}[h]{ccccccccccc}
\hline \hline
Time    & Observations & Disc Norm.$^b$       & $kT_{\rm in}$ & $kT_{\rm e}$ (keV) & $\tau^c$ & $E_{\rm Fe}$ line    & $\omega^d/2\pi$& \kir             & $F^e$                      & $F_{\rm bol}^f$\\
(MJD$^a$)   &    (Rev. \#)          &                      & (keV)         &   or $\Gamma$ &       & (keV)  &              & (dof)            & & \\
\hline

217.7--217.8 & N.A.$^g$ & 33784$_{17644}^{+28803}$  & 0.18$\pm$0.01 & $\Gamma$=1.62$\pm$0.02 & - & 6.53$_{-0.35}^{+0.32}$ & 0.34$_{-0.08}^{+0.10}$ & 1.45 (281) & 1.43 & N.A. \\

237.5--240.2 & 895$^h$ & 35$_{-9}^{+20}$ & 0.83$\pm$0.10  & 41$_{-4}^{+5}$ & 1.63$_{-0.17}^{+0.14}$  & 6.56$_{-0.25}^{+0.20}$ & 0.25$\pm$0.05 & 0.84 (97) & 1.99 & 0.82\\ 

240.7--243.2 & 896$^h$ & 100$_{-50}^{+186}$ & 0.67$_{-0.11}^{+0.02}$ & 38$_{-3}^{+4}$ & 1.69$_{-0.14}^{+0.13}$ & 6.15$_{-0.15}^{+0.21}$ & 0.19$_{-0.04}^{+0.05}$ & 1.05 (102) & 2.19 & 0.89 \\
 
243.7--246.2 & 897$^h$ & 100$\pm$50 & 0.65$\pm$0.10 & 41$_{-3}^{+5}$  & 1.53$_{-0.15}^{+0.12}$  & 6.00${_i}^{+0.14}$ & 0.21$\pm$0.03 & 1.13 (103) & 2.14 & 0.86\\

246.7--249.1 & 898$^h$ & 100$_{-50}^{+91}$ & 0.64$_{-0.07}^{+0.03}$ & 39$_{-3}^{+6}$ & 1.59$_{-0.15}^{+0.14}$ & 6.00${_i}^{+0.25}$  & 0.22$\pm$0.05 & 1.14 (102) & 2.29  & 0.92 \\

249.7--252.1 & 899$^h$ & 66$_{-29}^{+252}$ & 0.75$_{-0.20}^{+0.14}$ & 40$_{-3}^{+5}$ & 1.59$_{-0.15}^{+0.13}$  & 6.37$\pm$0.20 & 0.26$\pm$0.05 & 1.09 (102) & 2.49 & 1.02 \\

252.5--255.1 & 900$^h$ & 102$\pm$51 & 0.68$_{-0.19}^{+0.13}$ & 39$_{-3}^{+4}$  & 1.59$_{-0.15}^{+0.13}$  & 6.17$_{-0.49}^{+0.23}$  & 0.25$_{-0.04}^{+0.05}$  & 0.78 (102) & 2.74 & 1.09 \\

255.6--258.1 & 901$^h$ & 124$_{-56}^{+388}$ & 0.70$_{-0.16}^{+0.11}$ & 45$_{-4}^{+7}$  & 1.36$_{-0.17}^{+0.14}$  & 6.31$_{-0.70}^{+0.30}$ & 0.32$\pm$0.04 & 0.99 (102) & 3.18 & 1.25\\
 
259.9--261.1 & 902 &  82887$_{-38606}^{+63704}$  & 0.19$\pm$0.01 &  38$_{-2}^{+3}$   & 1.49$_{-0.11}^{+0.10}$  & 6.23$_{-0.23}^{+0.15}$ & 0.25$\pm$0.04 & 1.58 (333) & 3.38 & 1.64 \\ 

264.5--267.1 & 904$^h$ & 134$_{-57}^{+441}$  & 0.73$_{-0.18}^{+0.12}$ & 39$_{-3}^{+4}$  & 1.45$_{-0.13}^{+0.11}$  & 6.23$_{-0.66}^{+0.29}$  & 0.34$_{-0.04}^{+0.05}$  & 1.07 (102) & 4.21 & 1.51 \\
 
267.6--270.1 & 905$^h$ & 183$_{-87}^{+1406}$  & 0.67$_{-0.20}^{+0.12}$ & 41$_{-3}^{+5}$  & 1.31$_{-0.14}^{+0.12}$  & 6.27$_{-0.70}^{+0.27}$  & 0.35$\pm$0.04 & 1.11 (102) & 4.60 & 1.61 \\ 

270.4--272.8 & 906$^h$ & 370$_{-199}^{+1664}$ & 0.60$_{-0.13}^{+0.11}$ & 38$_{-3}^{+4}$& 1.40$_{-0.12}^{+0.11}$ & 6.07$_{-0.47}^{+0.30}$ & 0.32$_{-0.03}^{+0.04}$ & 1.33 (102) & 5.04 & 1.77 \\ 

273.4--276.0 & 907$^h$ & 198$_{-71}^{+568}$ & 0.68$_{-0.15}^{+0.09}$ & 39$_{-3}^{+4}$ & 1.28$_{-0.13}^{+0.11}$ & 6.45$_{-0.50}^{+0.19}$ & 0.35$_{-0.04}^{+0.05}$ & 1.59 (102) & 5.67 & 1.98 \\
 
276.4--278.8 & 908$^h$ & 338$_{-174}^{+2463}$ & 0.62$_{-0.17}^{+0.13}$ & 37$_{-3}^{+4}$ & 1.28$_{-0.13}^{+0.11}$ & 6.29$_{-0.65}^{+0.35}$ & 0.39$\pm$0.06 & 1.15 (102) & 6.59 & 2.15\\

279.4--281.5 & 909 &  32816$_{-19118}^{+33761}$ & 0.21$_{-0.01}^{+0.02}$  & 39$_{-3}^{+4}$  & 1.16$_{-0.13}^{+0.11}$  & 6.23$_{-0.25}^{+0.15}$ & 0.39$\pm$0.04 & 0.88 (312) & 7.80 & 2.79 \\ 

281.7--281.8 & N.A.$^g$ &  52$_{-20}^{+24}$ & 0.85$\pm$0.09 & $\Gamma$=1.80$\pm$0.01 & - & 6.27$_{-0.22}^{+0.16}$ & 0.38$\pm$0.06 & 0.57 (255) & 7.94 & N.A. \\ 

282.4--284.9 & 910$^h$ & 32$_{-6}^{+23}$ & 1.19$_{-0.11}^{+0.10}$ & 40$_{-3}^{+4}$ & 1.14$\pm$0.12 & 6.00${_i}^{+0.58}$  & 0.49$\pm$0.04 & 0.96 (96) & 7.69 & 2.25\\ 

289.3--289.6 & N.A.$^g$ & 840436$_{-505374}^{+118692}$  & 0.16$_{-0.01}^{+0.02}$ & $\Gamma$=1.79$\pm$0.03 & - & $6.14_{-0.29}^{+0.21}$ & 0.28$\pm$0.07 &  1.34 (228) & 9.65 & N.A. \\ 

297.8--297.9 & N.A.$^g$ & 174096$_{-62284}^{+86620}$ & 0.22$\pm$0.01 & $\Gamma$=1.95$\pm$0.02 & - & 6.00${_i}^{+0.04}$ & 0.10$\pm$0.05 & 1.32 (260) & 11.4 & N.A. \\ 

301.1--303.3 & 916 & 3196$_{-288}^{+367}$  & 0.61$\pm$0.02  &  256$_{-83}^{+6}$  & 0.01$_{-0.01}^{+0.02}$ & 6.00$_{i}^{+0.22}$ & 0.51$_{-0.10}^{+0.09}$ & 1.91 (252) & 12.2 & 2.93 \\

303.6--303.8 & N.A.$^g$ & 2652$_{-353}^{+526}$ & 0.73$\pm$0.04 & 35$_{-14}^{+5}$  & 0.50$_{-0.50}^{+0.28}$ & 6.74$_{-0.14}^{+0.16}$ & 0.36$_{-0.32}^{+0.44}$ & 1.77 (192) & 14.6 & N.A.\\

310.6--310.7 & N.A.$^g$ & 999$_{-44}^{+45}$ & 1.00$_{-0.01}^{+0.02}$  &  $\Gamma$=2.33$_{-0.07}^{+0.06}$ & - & 6.15$_{-0.09}^{+0.08}$ & - & 1.50 (178) & 13.3 & N.A. \\ 

\hline
\end{tabular}
\end{center}
Notes:\\
Models applied in {\tt XSPEC} notations: {\sc constant}*{\sc
phabs}*{\sc reflect}*({\sc diskbb}+{\sc gaussian}+{\sc powerlaw}) or  {\sc constant}*{\sc
phabs}*{\sc reflect}*({\sc diskbb}+{\sc gaussian}+{\sc comptt}) (not all components always needed). $\Gamma$ is the photon index. $N_{\rm H}$ varied between 4.9--6.3~$\times$~10$^{21}$ cm$^{-2}$. Errors are given at the 90\% confidence level ($\Delta \chi^2=2.7$).\\
~a) MJD-55000.\\
~b) Disc normalization $K$ is proportional to $(R/D)^{2}\cos\theta$, where $R$ is the
inner disc radius in km, $D$ is the distance to the source in kpc and $\theta$
the inclination angle of the disc.\\
~c) Plasma optical depth.\\
~d) Reflection scaling factor (1 for isotropic source above disc).\\
~e) Computed in the 2--20~keV range. Units: 10$^{-9}$~erg~cm$^{-2}$~s$^{-1}$. \\
~f) X-ray bands extrapolated in the 0.01~keV--10~MeV range. Units: 10$^{-8}$~erg~cm$^{-2}$~s$^{-1}$.\\
~g) N.A. = Non Applicable (only XRT and PCA data available).\\
~h) XRT data not available.\\
~i) Pegged at lower limit (6 keV).\\
\end{table*}

\begin{figure}
\centering
\includegraphics[width=6.3cm, angle=270]{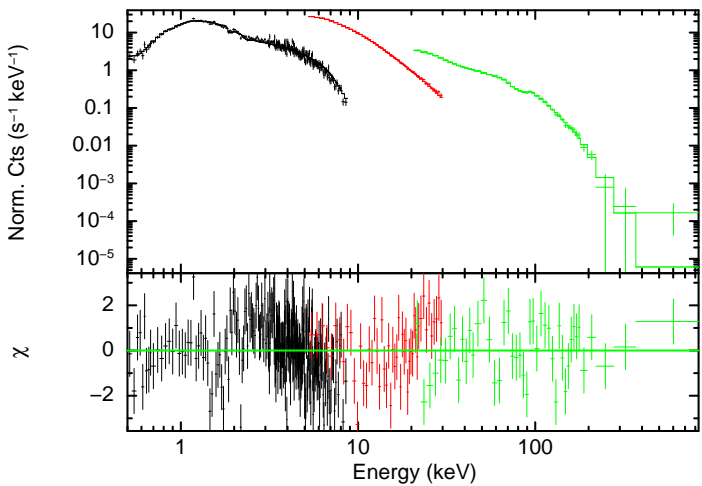}
\caption{\label{spe1}~{\emph{Swift}}/XRT (black), {\emph{RXTE}}/PCA (red) and {\emph{INTEGRAL}}/ISGRI (green) LHS count rate spectra of GX 339-4 (first ToO of March 2010, Rev. \# 902, $\sim$55259.9--55261.1 MJD) fitted with an absorbed multi-colour disc, reflection, Fe line and Comptonization components.}
\end{figure}

\begin{figure}
\centering
\includegraphics[width=6.3cm, angle=270]{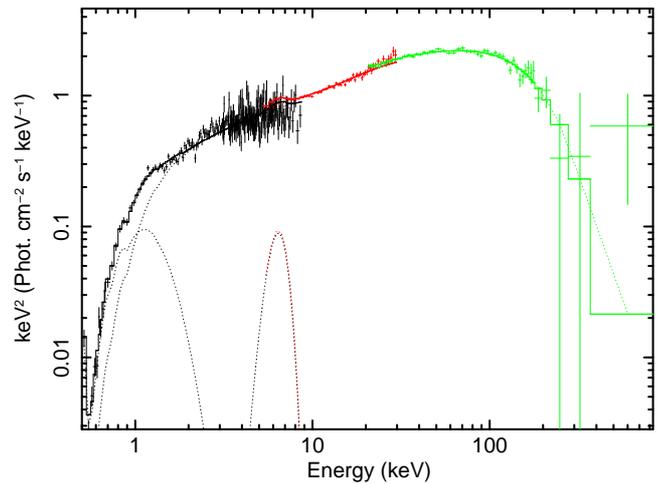}
\caption{\label{spe2}~Same as Fig.~\ref{spe1}, but in distinct units: keV$^2$~$\times$~Photons cm$^{-2}$ s$^{-1}$ keV$^{-1}$.}
\end{figure}

\begin{figure}
\centering
\includegraphics[width=9.2cm]{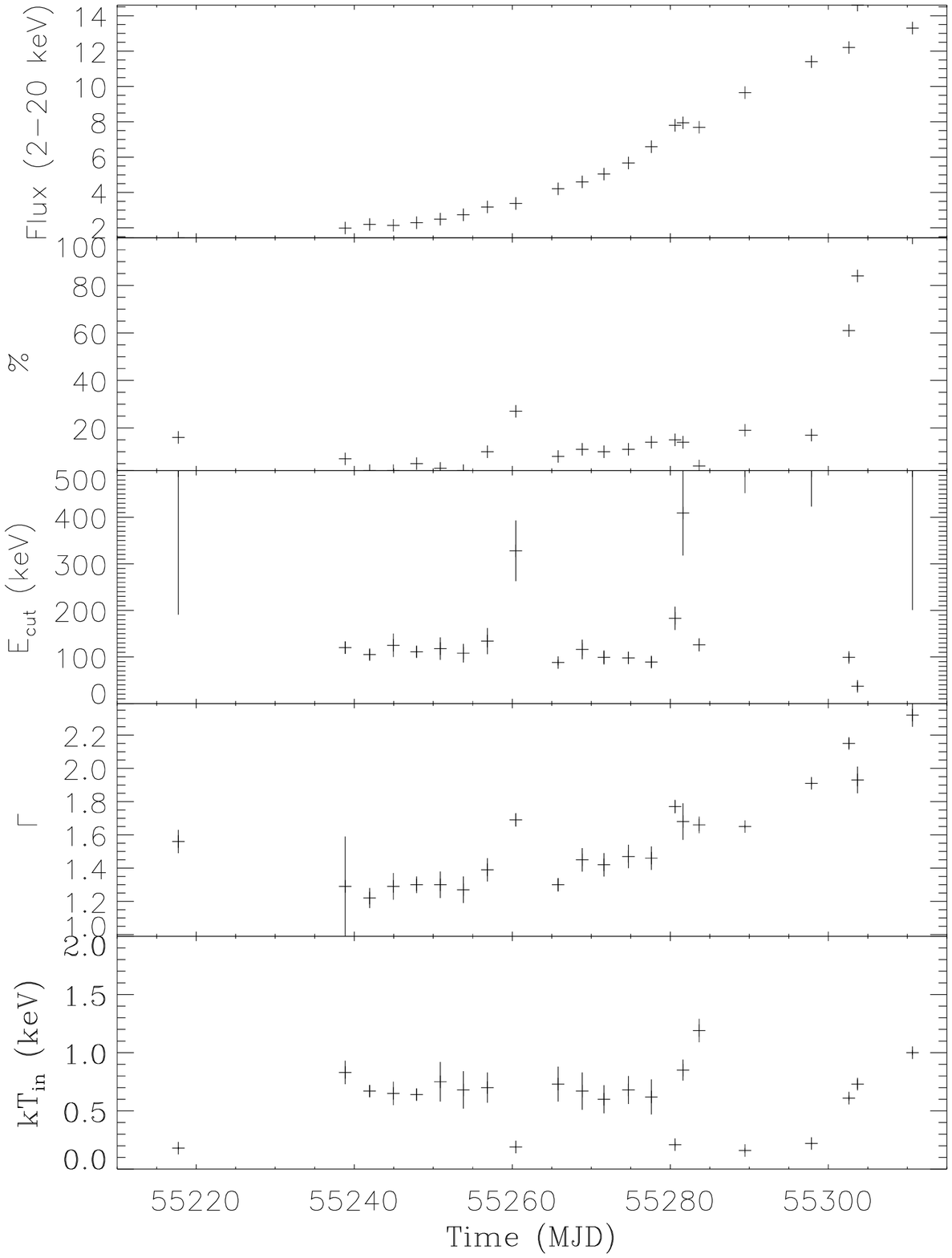}
\caption{\label{param-boff}~Summary of the parameter evolutions of \gx\ along its outburst: 2--20 keV flux (10$^{-9}$~erg~cm$^{-2}$~s$^{-1}$), \% of disc versus power law flux, high energy cut-off energy (keV), corresponding photon index and inner disc temperature (keV). See Tables~\ref{tab:para} and \ref{tab:cutoffPL}, and text for details.}
\end{figure}

We fitted the closest (in time) XRT (0.5--9.2~keV, when available), PCA (3--30~keV) 
and IBIS/ISGRI (18--200~keV) spectra simultaneously with {\tt XSPEC v12.6.0} \citep{Arnaud:1996}. 
Several models were tested when analyzing the spectra. A normalization constant was added to account for 
uncertainties in the cross-calibration of the instruments. 
The data were in general well fitted using absorbed 
power law (or cut-off power law and Comptonization) with reflection combined with a multicolour
black-body, and a Gaussian at $\sim6.5$~keV attributed to a fluorescent iron (Fe) line. The line centroid
could not be well constrained by the PCA: we thus fixed the energy of the line to lie between $\sim$ 6~keV and 7.5 keV. 
An additional line at 2 keV (due to calibration problems) was sometimes needed in XRT. Absorption was a combination of 
a fixed interstellar component (abundances fixed to \citealt{wilms00}) plus a variable local component. XRT found the total 
absorption to vary between 4.9--6.3~$\times$~10$^{21}$ cm$^{-2}$, taking into account 
the statistical uncertainties ($\pm$0.07), and the dependence on the position 
of the source in the HID. The lower end (4.9~$\times$~10$^{21}$ cm$^{-2}$) 
is compatible, within the errors, with the average Galactic
 column density in the source direction estimated from \citet{kalberla05}; it is possible that, up to about 1.4~$\times$~10$^{21}$ cm$^{-2}$ is intrinsic and variable 
 (see \citet{cabanac} for a detailed discussion of the variations of the absorption in this source; XRT data were not always 
 available to do that in this work). For the disc component, 
the {\sc diskbb} model in {\tt XSPEC} \citep{Mitsuda:1984} was used. We then replaced the phenomenological models, 
that first allowed us to easily compare the spectral parameters over the outburst, 
with a more physical one, the thermal Comptonisation model of \citet{Titarchuk:1994}. We 
tested the best modelling by adding additional contributions and 
we carefully checked any improvement in the goodness of the fit. 
When a cut-off was needed, this provided a good fit to almost all our 
data (see Table~\ref{tab:para} and \ref{tab:cutoffPL} for details; otherwise, we used a simple 
power law model), although the temperature of the seed photons (tied to the disc temperature) was not always well constrained. \\
\noindent The best-fit parameters are reported in Table~\ref{tab:para}; Table~\ref{tab:cutoffPL} shows the 
phenomenological cut-off power law parameters. 
In addition to providing a more physical interpretation to the data, this model also permits us to avoid 
the divergence of the power~law flux towards low energy since Comptonisation significantly 
contributes above 3$\times$ $kT_{\rm in}$, and decreases significantly below this energy. Our spectral fitting 
in general lead to normalization constants very close to 1 for all instruments. 
Examples of the fitted spectra obtained during our \integral\ observations are shown 
in counts in Fig.~\ref{spe1} with residuals, and in Fig.~\ref{spe2} in energy units (keV$^2$ $\times$ Photons 
cm$^{-2}$ s$^{-1}$ keV$^{-1}$). A small excess was seen at high 
energies, but a non-thermal modelling did not improve our fits: this excess was not significant. 
A plot summarizing the main spectral parameters for all the observations 
is shown in Fig.~\ref{param-boff}. Large error bars correspond to poorly constrained data.\\
\begin{table*}[t!]
\begin{center}
\caption{\label{tab:cutoffPL} Cut-off power law best-fit parameters for the periods reported in Table~\ref{tab:para}.}
\begin{tabular}{cccccc}
\hline \hline
Time    & \integral &  $\Gamma$ & $E_{\rm c}$            & \kir & $P^b$                      \\
(MJD$^a$)   &    Observations (Rev. \#)                   &        & (keV)             &  (dof) &\\
\hline

217.7--217.8 & N.A. & 1.56$_{-0.05}^{+0.08}$& 500$_{-309}^{c}$ & 1.56 (280) & 16 \\

237.5--240.2 & 895 & 1.29$_{-0.13}^{+0.47}$ & 120$_{-12}^{+15}$ & 0.84 (97) & 7\\ 

240.7--243.2 & 896 & 1.22$\pm$0.06 & 105$_{-10}^{+12}$ & 1.06 (102) & 2 \\
 
243.7--246.2 & 897 &1.29$_{-0.08}^{+0.09}$ & 125$_{-22}^{+34}$ & 0.75 (103) & 2 \\

246.7--249.1 & 898 & 1.30$_{-0.06}^{+0.05}$ & 111$_{-10}^{+14}$ & 1.18 (102) & 5 \\

249.7--252.1 & 899 &1.30$\pm$0.08 & 118$_{-18}^{+30}$ & 1.23 (102) & 3 \\

252.5--255.1 & 900 & 1.27$\pm$0.08 & 108$_{-16}^{+25}$ & 0.77 (102 ) & 2 \\

255.6--258.1 & 901  & 1.39$\pm$0.07& 134 $_{-23}^{+33}$& 1.02 (102) & 10 \\

259.9--261.1 & 902  & 1.69$\pm$0.04 & 328$_{-64}^{+94}$ & 1.69 (333) & 27 \\

264.5--267.1 & 904 & 1.30$\pm$0.04 & 88$\pm$6 & 1.21 (102) & 8\\

267.6--270.1 & 905 & 1.45$\pm$0.07 & 116$_{-17}^{24}$& 1.21 (102) & 11 \\

270.4--272.8 & 906 & 1.42$\pm$0.07 & 99 $_{-13}^{+17}$ & 1.49 (102) & 10 \\

273.4--276.0 & 907 & 1.47$_{-0.06}^{+0.07}$ & 98$_{-11}^{+15}$ & 1.66 (102) & 11 \\

276.4--278.8 & 908 & 1.46$_{-0.06}^{+0.07}$ & 89$_{-10}^{+14}$ & 1.17 (102) & 14 \\

279.4--281.5 & 909 & 1.77$\pm$0.04 & 183$_{-23}^{+28}$ & 0.92 (312) & 15  \\

281.7--281.8 & N.A. & 1.68$_{-0.11}^{+0.10}$ & 409$_{-247}^{c}$ & 1.49 (254) & 14 \\

282.4--284.9 & 910 & 1.66$_{-0.04}^{+0.05}$ & 126$_{-10}^{+17}$ & 1.53 (96) & 4 \\

289.3--289.6 & N.A. & 1.65$\pm$0.01 & 500$_{-48}^{c}$ & 1.73 (227) & 19 \\

297.8--297.9 & N.A. & 1.91$\pm$0.01 & 500 $_{-77}^{c}$ & 1.86 (259) & 17 \\

301.1--303.3 & 916 & 2.15$\pm$0.03 & 99  $_{-7}^{9}$ & 1.69 (252) & 61 \\

303.6--303.8 & N.A. & 1.93$_{-0.09}^{+0.08}$ & 37$_{-7}^{+10}$ & 1.65 (192) &  84 \\

310.6--310.7 & N.A. & 2.32$_{-0.07}^{0.06}$ & 500 $_{-299}^{c}$ & 1.55 (177) & 100 \\
\hline
\end{tabular}
\end{center}
Notes:\\
~a) MJD-55000.\\
~b) Percentage of the disc flux versus the power law flux.\\
~c) Pegged at hard limit (500 keV).\\
\end{table*}
\noindent Apart from three very low values around 55217, 55260 and 55280 MJDs, the disc temperature 
values were remarkably constant, around 0.6--0.8 keV, until the source transited to a softer state. 
After the transition, the disc temperature was first found at its lowest values, but it returned to its pre-transition value 
(by 55302 MJD) within a time frame of about 10--12 days, even though the power law index 
continued to increase (see next paragraph). The observed inner disc temperature
and the normalization evolution which could be high (up to $\sim$ 55303 MJD, Rev. \#916) before decreasing 
(see Table~\ref{tab:para}), correspond to what is usually 
observed for \gx\ in outburst (see Sect.~\ref{inter}).
The radius inferred from {\sc {diskbb}} is however questionable \citep{Gierlinski:2008}. 
Several corrections need to be applied (colour correction factor, irradiation of the disc distorting
the spectrum, physical prescription for the inner boundary condition),
but these have large uncertainties. Moreover, the size of the correction
can vary with time. Therefore, in Fig.~\ref{param-boff} we only show 
the fraction of the flux that is in the {\sc{diskbb}} component with respect to the power law flux: as expected, 
its contribution increased with time. 
The reflection component also evolved accordingly: its contribution 
increased with time, from 0.19 to 0.51, compatible with the 
source going to a softer state.

\noindent From 55217 to 55311 MJD, the power law photon index varied between 1.6 and
2.4 (or a cut-off power law between 1.3--2.3, Table~\ref{tab:cutoffPL}). 
In the early stages of the outburst, a cut-off is needed in the spectral fits (the Comptonization model 
improves the \kir), whereas it 
disappears in the softer states. Indeed, 
later data are better fit with a simple power law model (Table~\ref{tab:para}). 
This indicates a cooling of the corona as the thermal component gets stronger, 
and then a change of the emission mechanism at the state transition. 
This further confirms that the source made a transition from a hard to soft state. With the model of \cite{Titarchuk:1994}, 
we obtained temperatures ($kT_{\rm e}$) and optical depth ($\tau$) values before Rev. \# 916
that yield a Comptonization parameter (y$\times$$m_{\rm{e}}c^2$) of 0.10--0.22, typical for a BH in the rising phase of the LHS. 
The bolometric flux extrapolated between 0.01 keV and 10 MeV (note that the bulk of emission is between 0.1 and 100 keV) 
rose from 0.8 to 2.9$\times$ 10$^{-8}$~erg~cm$^{-2}$~s$^{-1}$ while the 2--20 keV flux rose from 1.4 to 14.6 $\times$ 
10$^{-9}$~erg~cm$^{-2}$~s$^{-1}$. After that, the spectral parameters were 
compatible with the source making a transition from the LHS to the HIMS (until $\sim$ 55303.6 MJD), then to softer states 
(to the SIMS until 55304.7 MJD, then quickly to the SS, $\sim$ 55306.1 MJD, and finally 
back to the SIMS and HIMS (see Fig.~\ref{hid}) around 55320 MJD \citep{Motta10b}, 
but this is beyond the scope of our paper since it happened after the observations listed in our Tables. 

\subsection{Multi-wavelength studies}

Observations at other wavelengths of \gx\ were conducted with {\it Swift}/UVOT from $\sim$ 55280 to 55300 MJD \citep{yu10}, aimed at fast optical/UV variability in the hard state and transition. No rapid optical/UV variability was detected due to the low UVOT count rates. 
Our long-term FTS optical light curves of this outburst are shown in Fig.~\ref{lcall}. 
The source brightened by $\sim 4.5$ mag in $V$, $R$ and $i'$-bands from a low luminosity state in 
2009 October, 55126 MJD  \citep[]{lew09}. \gx\ reached the brightest magnitudes recorded, 
on 2010 April 1, 55287 MJD  ($V$ = $15.05 \pm 0.02$; $R$ = $14.49 \pm 
0.01$; $i'$ = $14.26 \pm 0.01$), just before the start of the transition from the LHS to softer states. 
During this LHS rise, the source brightened at a mean rate of $\sim 0.01$ mag d$^{-1}$ 
between 55237--55287 MJD. 
There was considerable optical short-term variability in the LHS. From several consecutive exposures between 55273--55320 MJD, we measured a fractional rms variability in the $i'$ band that was fairly constant: $\sim 9$--11\% with typical errors of $\sim 1$ \% (on each date, the time resolution was between 90 and 130 seconds). This short-term variability throughout the outburst, including the decay, will be detailed in another paper (Russell et al. 2011, in prep.). Our results are similar to the optical fractional rms previously reported at time resolutions of $< 1$ sec in the LHS of \gx\ \citep[see, e.g.,][]{gand10}. Between 55294--55297 MJD, the optical flux faded rapidly by $\sim 1$ mag in 3.0 days \citep{Ru10} and a change in the SED to a bluer colour was seen (see Sect.~\ref{sed-discuss}). 
At the same time the {\it Swift}/BAT flux started to fade, marking the beginning of the 
transition away from the LHS (Fig.~\ref{lcall}). The rapid optical fading was also accompanied 
by a drop in the $i'$-band fractional rms variability to $1.3 \pm 0.7$ \% by $\sim$55301 MJD. 
As the source continued to soften in X-rays (Fig.~\ref{hid}), the optical flux remained fairly constant in the following few months, around $V$ $\sim 16.5$; $R$ $\sim 16.0$ and $i'$ $\sim 16.0$ mag, with an i'-band fractional rms of $< 5$ \%. 
REM data showed that the source was basically at a
constant luminosity on 55260-55261 MJD (during Rev. \# 902)
while on 55303 MJD (during Rev. \# 916) the source faded
consistently by a factor of $\sim 2$ in the optical ($V$ and $R-$band) and
by a factor of $\geq 5$ in the NIR ($I$ and $H-$band), in agreement with the FTS results. 
Months later, after the peak of the outburst, the optical and NIR brightnesses 
in the $V$, $R$, $I$, $J$, $H$ and $K$ filters all went down. A spectrum was taken with the
ISAAC instrument on the ESO telescope between 55261.3--55261.4 MJD and 
55307.2--55307.2 MJD, and had a flux and spectral shape similar to our REM 
and FTS photometric data (F. Rahoui et al. 2011, in prep.).

\noindent The ATCA observations conducted on 55283 MJD showed 
flux densities $\sim$ 20 mJy with an inverted radio spectrum, with a 
spectral index in the range: +0.1 to +0.2, typical of powerful self-absorbed compact jets observed in the LHS. 
On 55372 MJD, no radio emission was detected at the location of GX 339$-$4: this was consistent with 
the source being in a soft state \citep{corb10b}. 
Ejecta very close to the core of the system were detected with a spectral index of -0.34$\pm$0.08 
(which is typical of optically thin synchrotron emission). The new radio source was in the same direction as the large scale jets already 
detected in \gx\ (see \citealt{Gallo04}): the impact of material ejected by the system into the ISM was observed (see Discussion Section).

\section{Discussion}
\label{discussion}

\subsection{Interpretations}
\label{inter}

According to \citet{Wu10}, this bright outburst should have peaked at $>$ 0.83 Crab in hard X-rays. 
This is based on the empirical relationship between the hard X-ray peak flux and the waiting time 
since the last bright ($>$ 0.12 Crab) outburst of \gx. The source transited at almost the expected time and flux (see below); 
the softening of the X-ray flux at this transition occurred at a slightly lower flux than 
the 2002--2003 and 2007 outbursts, but at a brighter flux than the 2004--2005 outburst. 
The transition may have occurred slightly earlier as there 
were a few low-level LHS mini-outbursts between the 2007 and 
2010 outbursts \citep{Kong08,Mark09}. These mini-outbursts could have  
emptied the disc a bit more than if the source was in true quiescence the whole time. The BAT, ISGRI, PCA and ASM light curves 
show that the hard X-rays were dominated by the power law component while the soft X-rays were 
mainly due to the disc (whose flux increased with time) plus the power law (which was fading at the same time). 
When the power law component returns, we observe the spectral transition back from soft into harder states, 
and a secondary peak in the hard X-rays. The outburst evolution of \gx\ during the 2010 episode is fairly typical of BH outbursts in general. \\

\noindent Our spectral analysis showed that the source transited from an initial LHS, with a spectrum 
dominated by Comptonization in the beginning of the outburst ($\sim$ 55217-- 55281 MJD), 
to softer states where the power law cut-off was not needed anymore 
and the photon index of the power law was very soft ($\sim$ 2.3). 
In X-ray binaries, a cut-off power law spectrum is usually interpreted as the signature of inverse 
Comptonisation of soft seed photons by a thermalised (i.e., with velocities following a 
Maxwellian distribution) population of electrons. Changes in the hard component 
can signal the presence of a compact jet, a corona, or reprocessed hard X-ray emission 
due to X-ray heating from an extended central source. The hard component therefore evolved as expected. 
Regardless of the caveats mentioned in Sect.~\ref{xgamspec} about the {\sc{diskbb}} components, and the exact value of the inner 
radius obtained from the fits, we see a trend in the disc parameters (Table~\ref{tab:para} and Fig.~\ref{param-boff}) which 
is compatible with an increase of the mass accretion rate before the transition. Then 
the disc globally recedes when the source reaches a low luminosity \citep[see, e.g.,][]{tomsick09}, 
during the decay phase of the outburst. Although we have then observed the disc slowly moving 
outwards during the hardening \citep{Chen:1997,CadolleBel:2004}, 
this is not necessarily true for all transient sources (see, e.g., \citealt{Miller:2006}). This is 
still strongly debated \citep{Done:2007,Rykoff:2007,Gierlinski:2008} 
as, for example, in XTE J1817$-$330. We do not address this
question specifically in this work as the majority of our data were taken in the LHS and not long after
the main transition into softer states. In addition, with a lower energy limit often at 3 keV, 
few data are sensitive enough for this purpose. However, 
one can remark that the rather low disc temperatures are consistent 
with the compact object being a BH and not a neutron star \citep[e.g.,][]{Tanaka:1995}. 
The soft and hard X-rays evolved as seen in previous outbursts of \gx: first in the HIMS, 
the source transited to softer states where the disc dominated the emission, the inner disc moved closer to the BH 
and then gradually cooled down before returning to quiescence.

\noindent The spectral evolution is consistent with other works: \rxte\ monitoring showed softening and LFQPO evolutions 
\citep{shap10,motta10,yu10} as \gx\ started to leave the LHS \citep{motta10}. Fig.~\ref{param-boff} shows some differences with 
the Fig. 6 of \citet{Motta09}: while the fluxes, disc temperature and power law indices show the same 
trend, our individual cut-off values are sometimes very high 
(e. g., near 55260 MJD), like in the HSS of their Fig. 6, and 
no cut-off is needed. 
The evolution in the rising phase of the LHS is therefore not as smooth as expected at higher 
energies (which are well constrained for the first time), but there is no clear explanation for that. 
The 2--20 keV flux (from 1.43 to 14.6 $\times$ 
10$^{-9}$ erg cm$^{-2}$ s$^{-1}$) and the bolometric flux (extrapolated from 0.01 to 1000 keV) variations 
(0.82 to 2.93$\times$10$^{-8}$ erg cm$^{-2}$ s$^{-1}$) confirm this spectral evolution. 
The source reached a high luminosity at the maximum of $\sim$ 12.9 $\times$ (d/6 kpc)$^2$ 10$^{37}$ erg s$^{-1}$. This 
represents $\sim$ 18\% of the Eddington luminosity for a 7 solar mass BH; stellar mass BHs accreting at or 
below 10$^{-2}$ $L_{\rm Edd}$ are found in the LHS \citep{McClintock:2006}. Besides, 
the evolution of the reflection component from 0.19 to 0.51 
indicates that the source became softer. Later than the observations 
presented in this paper, the source switched back to the SIMS, and subsequently to the 
HIMS. This is not unusual: during the decay of their outbursts, 
XTE J1720$-$318, XTE J1650$-$500 and SWIFT J1753.5$-$0127 showed a slowly receding disc with 
a decreasing inner temperature, while at the same time the relative amount of the power law contribution increased again. 
A year after the beginning of its outburst (end of February 2011), \gx\ finally switched back to the 
LHS \citep{Russ11c}, and then to quiescence \citep{Russ11d}. \\

\noindent The broadband spectral parameter evolution from hard to soft states, as well as our radio/NIR/optical studies,
are consistent with the preliminary results of \citet[]{Lew10,corb10a}. A discrete radio ejection usually 
occurs around the hard to soft transition in a BH transient outburst: 
\cite{corb10b} witnessed with the ATCA the interaction of a relativistic jet from \gx\ 
with the interstellar medium, implying that a major ejection occurred earlier. During the LHS, 
we detected the compact core jet in the optical/IR. The source continued to rise in the LHS for a while \citep{Wu10}, 
therefore it was also rising at radio frequencies. We added radio flux data to our SED  
(Fig.~\ref{sed1bisfit}) but the entire campaign will be presented 
in a forthcoming paper (Corbel et al. 2011, in prep.). The rapid drop in optical flux and colour change 
observed at the start of the state transition are 
reminiscent of previous outbursts \citep{coriat09}. Since rapid variability probably originates in the synchrotron jet component \citep{casella10,gand10}, this is consistent with the jet no longer making a strong contribution to the optical emission and that it was fading 
\citep{Ru10}. Such chromatic
behaviour, also observed in the REM data, hints for the presence of a component mainly contributing in the
NIR frequencies which then fades or disappears after the transition (see also Fig.~\ref{sed4}). 
The optical and infrared flux continued to drop in the days following the transition. The 
changes in optical flux and spectrum over the transition have been reported 
before for \gx\ \citep{Homan:2005b,coriat09} and other BH LMXBs \citep[see, e.g.,][]{bux04,russ07}. In this paper, 
we clearly detect the jet evolution and its dramatic quenching at all wavelengths.

\subsection{Spectral energy distributions}
\label{sed-discuss}

\begin{figure*}[t!]
\centering\includegraphics[angle=270,width=0.6\linewidth]{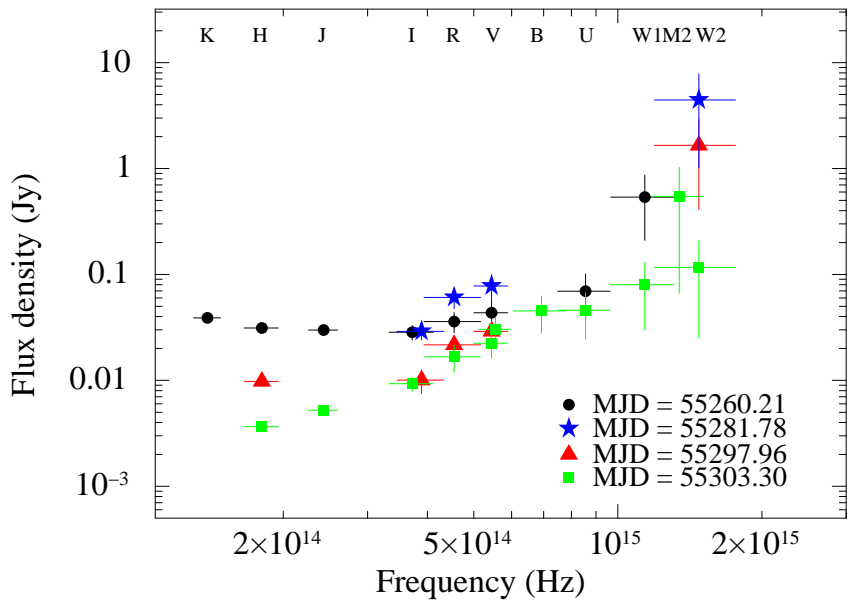}
\caption{\label{sed4}{Compiled SEDs with UVOT, REM, SMARTS (M. Buxton et al. 2011, A\&A, submitted) 
and FTS data at distinct MJD: 
MJD = 55260.21 (UVOT/$uw1$, $u$ and REM/$V$, $R$, $I$, $J$, $H$, $K$); MJD = 55281.78 (UVOT/$uw2$ and FTS/$V$, $R$, $i'$); 
MJD = 55297.96 (UVOT/$uw2$, FTS/$R$, $i'$ and SMARTS/$V$, $H$) and MJD = 55303.30 
(UVOT/$uw2$, $um2$, $uw1$, $u$, $b$, $v$, SMARTS/$V$, $I$, $J$, $H$ and REM/$R$). 
The error bar is so large that the fluxes of $uw1$, $um2$ and $uw2$ are comparable (within the errors).}}
\end{figure*}

To compute the NIR/optical spectral energy distributions (SEDs) of the source, we
first corrected our magnitudes for interstellar absorption. We 
assumed a colour excess of $E(B-V) = 1.2 \pm 0.1$ \citep{Zd98}, which is fully consistent with 
our measurements of $N_{\rm H}$ and of \citet{corbfen02}. 
Using standard extinction curves from \cite{Fi99} and \cite{Kata08}, we obtained 
the de-reddening parameters for each of our UV/optical/NIR filters (Table~\ref{tab:lambda_GX_339}). 
Figure~\ref{sed4} was obtained with the FTS, 
SMARTS and REM data at four distinct epochs (see caption). The spectra clearly changed 
between the FT South and SMARTS epochs ($\sim$55298 MJD).The jet faded over the transition and our 
last REM observations ($\sim$55303 MJD), while the other component (probably the disc) 
stayed about the same. The $H$-band faded by a factor of 10 whereas the $V$-band faded
 by a factor $>$2.5. The jet contribution moved to lower frequencies: 
 data taken on $\sim$ 55298 MJD still have the jet dominating 
in the $H$-band in NIR, but no longer in the optical. The 
last SED is bluer according to Fig.~\ref{sed4}. This is exactly what one can 
expect when a source undergoes a LHS to HSS spectral transition: 
the colour changed and thermal processes started to dominate. 
During the rising LHS, Gandhi et al. (2011, A\&A, submitted) found that the jet spectrum was highly variable in the mid-IR. The break between optically thick (self-absorbed) and optically thin synchrotron emission in the jet spectrum was found to vary 
between $\sim 3.6$ and 22 $\mu m$ on timescales of minutes--hours. Here, 
in Fig.~\ref{sed4}, we are witnessing this jet component fading over timescales of days.\\
\begin{table}
\caption{De-reddening parameters used to correct the UV/optical/NIR photometry of GX 339$-$4 assuming 
$E(B-V)=1.21 \pm 0.17$ mag, with the REM telescope, FTS and UVOT.}
\begin{tabular}{ccc}
\hline
Filter &${\lambda}_c$&  A$_{\lambda}$  \\ 
       &   (\AA)     &     (mag)       \\ 
\hline
REM- $V$    &  5505       & 3.64	    \\
$R$    &  6588       & 2.81	    \\
$I$    &  8060       & 2.04	    \\
$J$    &  12500      & 0.96	    \\
$H$    &  16500      & 0.62	    \\
$K$    &  21600      & 0.42	    \\
\hline
FTS-$V$ & 5448 & 3.69\\
$R$ & 6407 & 2.92 \\
$i'$ & 7545 & 2.27 \\
\hline
UVOT-$uw2$ & 1928 & 9.74 \\
$um2$ & 2246 & 10.97 \\
$uw1$ & 2600 & 7.91 \\
$u$ & 3465 & 5.93 \\
$b$ & 4392 & 4.88 \\
$v$ & 5468 & 3.67 \\
\hline
\end{tabular}
\label{tab:lambda_GX_339}
\end{table}
\begin{figure*}[t!]
\centering\includegraphics[angle=270,width=0.7\linewidth]{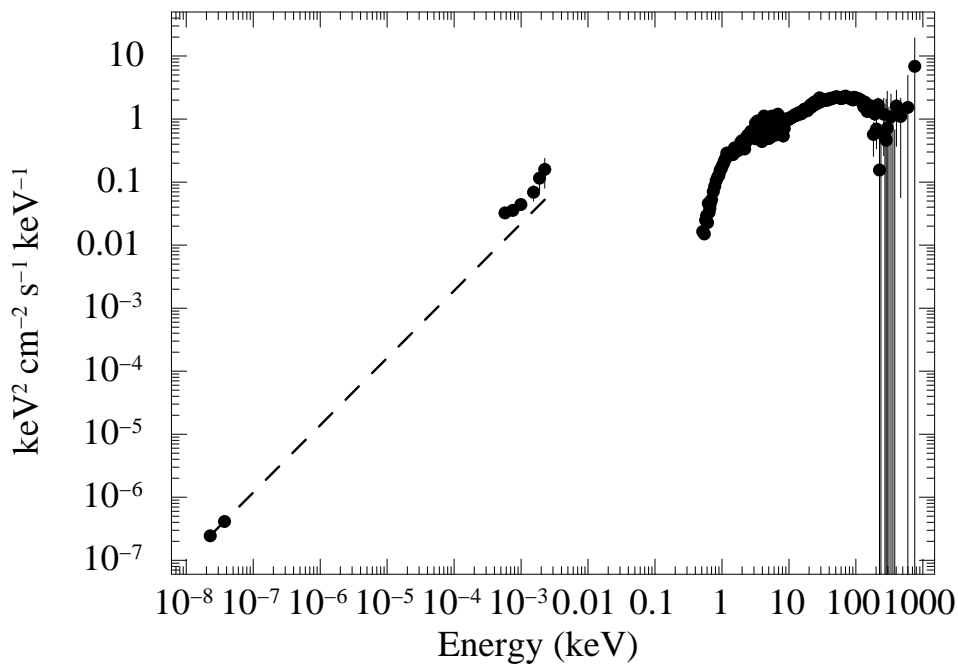}
\caption{\label{sed1bisfit}{SED from radio to soft $\gamma$-ray data plotted in flux (keV$^2$ $\times$ cm$^{-2}$ $\times$ s$^{-1}$ $\times$ keV$^{-1}$) versus energy (keV) at $\sim$ 55259.9--55261.1 MJD (Rev. \# 902). It is remarkably similar to the Fig.~2 of \cite{corbfen02}. 
The dash-line is a simple power law extrapolation of the two radio data points up to the NIR frequencies (this is not a physical model, but only shown for visual purpose; see Sect.~\ref{sed-discuss}).}} 
\end{figure*}
\begin{figure*}[t!]
\centering\includegraphics[width=0.7\linewidth]{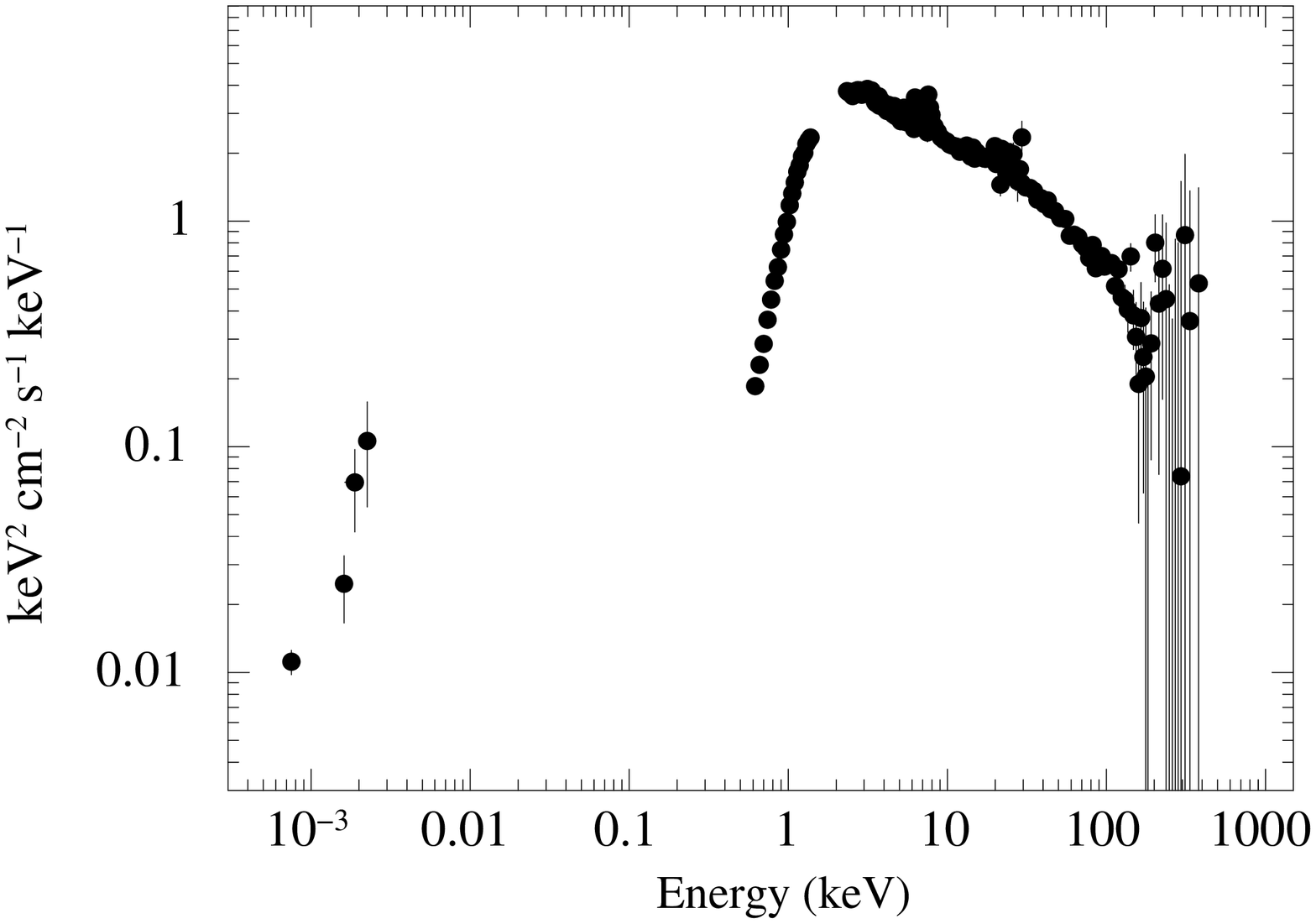}
\caption{\label{sed2}{SED from NIR to soft $\gamma$-ray data plotted in flux (keV$^2$ $\times$ cm$^{-2}$ $\times$ s$^{-1}$ $\times$ keV$^{-1}$) versus energy (keV)}
around 55301.1--55303.3 MJD (Rev. \# 916).}
\end{figure*}

\noindent We produced broader-band SEDs in two distinct spectral states (hard and soft), 
composed of simultaneous radio (when available), (de-reddened) UV/optical/NIR 
and unabsorbed X-ray/soft 
$\gamma$-ray data. 
They are shown in Figures~\ref{sed1bisfit} and \ref{sed2}. 
In Fig.~\ref{sed1bisfit}, we plot the data obtained during the first \integral\ ToO (55259.9--55261.1 MJD, Rev. \# 902). 
Fig.~\ref{sed1bisfit} is remarkably similar to the Fig. 2 of \cite{corbfen02}, focusing on GX 339$-$4 jet signatures during 
LHS. One can interpret the results with simple power laws, but note that this is a very rough approach and not a physical model. 
An extrapolation of the radio data up to the 
NIR/optical clearly does not fit the data. Excess NIR/optical emission is observed. 
Possible sources of the residual emission are the disc, 
the irradiated face of the companion star 
(as the companion cannot contribute much, see \citealt{Shab01,Hynes04}) and the jets. 
Similarly, several power law components with distinct slopes are needed to fit 
our SEDs.

\noindent For example, using a simple, double
power law fit to the data in Fig.~\ref{sed4} gives indices of $-0.9
\pm 0.4$ for the NIR range. We fixed the optical component (where the disc is assumed to dominate) to 
1.7 which is the value we found fitting a single power law to our SED around 55303 MJD,
where the jet contribution was absent or still negligible (thus assuming that,
at that epoch, we only have contribution from the disc). For this SED in
the soft state (Fig.~\ref{sed2}), we observed important spectral changes both in the
disc and hot medium components and found that a simple single power law
model was enough to fit the UV/optical/NIR data (Fig.~\ref{sed4}), with an index of
$1.7 \pm 0.2$. This is indicative of the fading jet component in \gx, as
previously discussed. The NIR/optical ESO/ISAAC spectroscopy taken at that time 
will be commented on in a forthcoming paper (Rahoui et al., 2011).\\
\begin{figure*}[t!]
\centering\includegraphics[width=0.6\linewidth,angle=270]{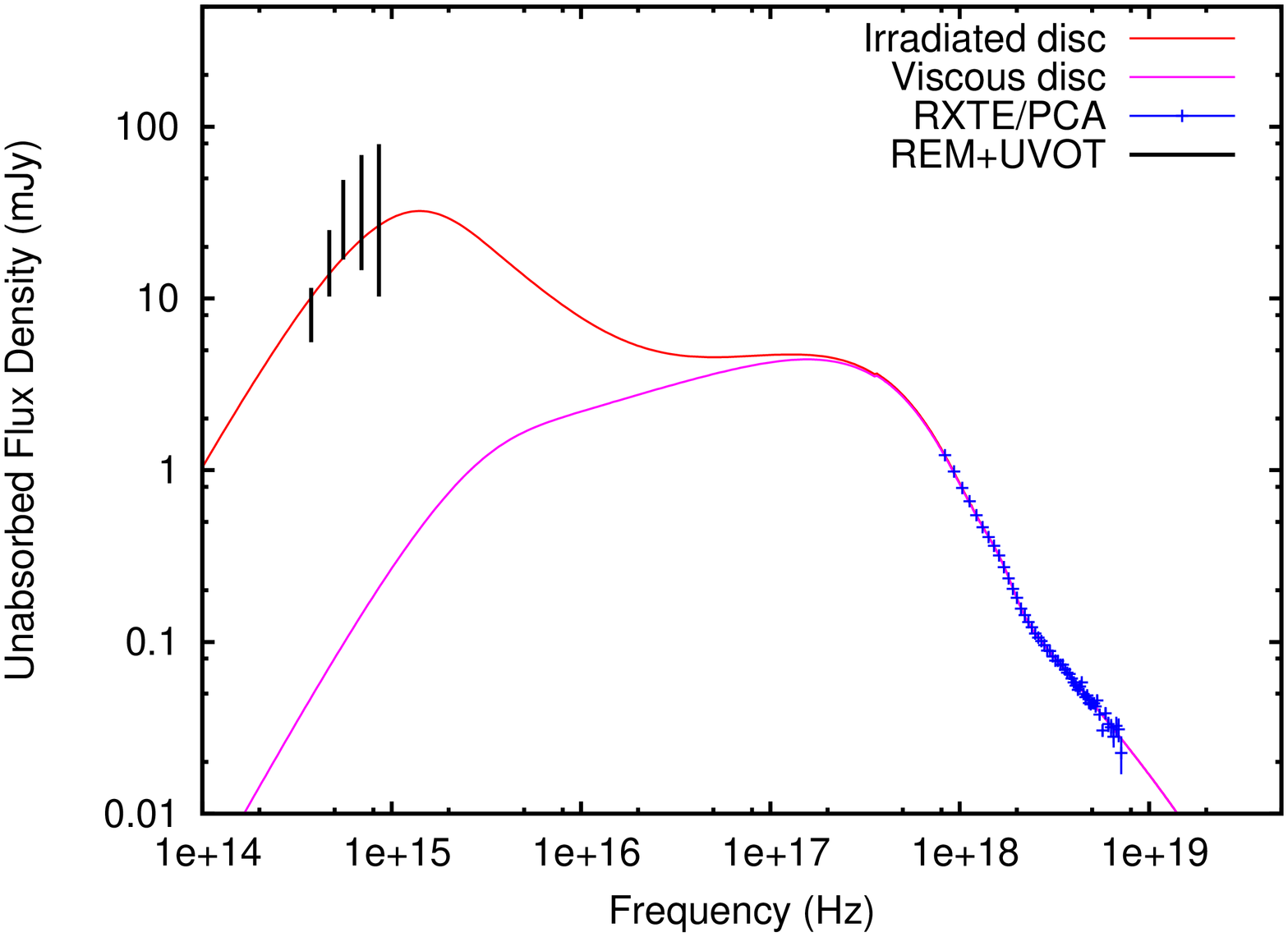}
\caption{\label{sedfarid}{Unabsorbed fluxes density (in mJy) versus frequency (Hz) from the 
RXTE (blue), REM and UVOT (black) data of \gx\ fitted with a Gaussian, a Comptonization model 
plus 1) an irradiated disc (red) or 2) a viscous disc (pink) 
around 55303.6--55303.8 MJD. The irradiated disc fits the data better (see Sect.~\ref{sed-discuss}).}}
\end{figure*}

\noindent After the LHS, the radio spectral index became typical of optically thin synchrotron radiation, probably as a result of freely
expanding plasma blobs previously ejected \citep[see, e.g.,][]{Fender:2004}.
This suggests that multiple ejection events took place during 
the outburst of \gx\ and then interacted with 
the interstellar medium \citet{corb10b}. 
This could potentially result in re-acceleration of particles up to very high energies. 
In general, the shape of our SEDs during the LHS are similar to the ones 
of the transient LMXB XTE J1118$+$480 \citep{Chaty:2003,Zurita:2006}: 
in its 2000 outburst, the SED from radio to X-rays has been explained as a combination 
of synchrotron radiation from a jet and a truncated optically thick disc, 
whereas models assuming advection dominated accretion flows alone underestimated the optical and IR fluxes 
(\citealt{Zurita:2006}, and references therein). In 2005, discrepancies observed between the optical and 
IR SEDs of XTE J1118$+$480 suggested that the IR was dominated possibly by a jet 
whereas the optical was dominated by disc emission. 
Power law fits to optical SEDs have also been 
performed for other BH XTs in outburst \citep{Hynes:2005}. All optical SEDs exhibit quasi power law spectra, 
with $\alpha$ ranging from 0.5--1.5, 
all steeper than that expected for a viscously heated, multi-temperature disc: 
$S_\nu\propto\nu^{1/3}$ (which differs from our value taken closer to the peak). 
Also, the authors found for these BHs that the UV/optical/X-ray data - when detected - could be fitted with a simple 
black-body model of an accretion disc heated by internal viscosity and X-ray irradiation, but the inner 
radius could not be well constrained. They concluded that the flat-spectrum synchrotron emission may 
be important in the IR and optical in this source. However, they did not exclude the alternative explanation 
that the IR excess could come from the cool outer disc. 
More recently, \citet{Russ2011} even showed in the colour-magnitude diagram of XTE J1550$-$564 
(Fig.~1 of their paper) that $\alpha$ can change dramatically over state transitions. In our observations of \gx\ presented here, 
the contribution of the radio to synchrotron emission up to the NIR/optical 
is important, and we saw it fading over the transition (see Sect.~\ref{inter}). However, another component, for example from the cooling 
disc and/or the irradiated companion, and/or an irradiated disc 
\citep{vanpar1994,Hynes:2005} might be necessary to account for the NIR/optical 
excess observed. Note that in quiescence, the NIR/optical of \gx\ is 
likely dominated by an optically thin disc plus a much fainter mass donor star \citep{Shab01}.\\

\noindent Finally, we had sufficient NIR and optical data to ascertain when \gx\ was in the soft state 
and constructed the corresponding SED around 55303.7 MJD (Fig.~\ref{sedfarid}). 
In fitting the SED, we used a technique similar to the one described in \citet[]{Rahoui10}: we tested a model 
consisting of a {\sc diskir} \citep{Gierlinski:2008,gier2009} added to a {\sc gaussian}, versus 
a model consisting of: {\sc diskbb} + {\sc gaussian} + {\sc comptt}. We fixed $kT_{\rm e}$ to the temperature found 
in Table~\ref{tab:para}. 
We found optical slopes around 2, which are 
consistent with thermal radiation. The reprocessing levels in the disc would need to be very high
to explain the observed UV and optical fluxes. Our results are consistent with \citet{coriat09} as they fit SEDs with a reprocessing model 
and found a peak-temperature in the UV, where the maximum irradiation is expected. 
However, we can not constrain the parameters very well (in fact, they are underestimated) with photometry alone.

\section{Conclusions}
\label{summary}

We obtained a unique data set, with the largest collection of multi-wavelength data 
collected so far simultaneously  for an outburst 
of \gx\ . We have presented light curves and spectral 
evolution of \gx\ in outburst from hard to soft states. We confirmed that 
the broadband behaviour was consistent with previous outbursts, 
though some peculiarities were seen at high energies. \cite{Hynes:2005} 
results on SEDs show similar behaviour for a collection of BHs (see also, 
e.g., \citealt{Russell:2006,russ07}). Extrapolating radio data to higher frequencies in the LHS SED
might be interpreted as non-thermal emission from 
ejected material, such as discrete ejection events that rapidly faded. Jet emission contributed significantly 
to the radio, NIR and optical data. An optical excess was seen above the jet 
component and could be explained by contributions from the disc (the companion 
can not account much for the emission as it is fainter than $R$ > 21 mag). 
However, alternative processes, such as 
X-ray irradiation, are sometimes inferred, depending on the source and 
data available. We now know that the amount of irradiation in NIR/optical varies depending on the 
luminosity and the X-ray state, as seen in our Fig.~\ref{sedfarid} around 55303.7 MJD. 
We did not observe the source switching back to the LHS 
as we did for XTE~J1720$-$318 \citep{CadolleBel:2004} or XTE~J1817$-$330
\citep{CadolleBel:2008}. For the first time, we have caught the jet quenching over the transition in many wavebands 
(Fig.~\ref{sed4}). This provides an observational constraint on how the jet evolved over the outburst. \\

\begin{acknowledgements}

We thank the referee for their helpful comments to improve our manuscript. 
We thank the \integral, \swift\ and \rxte\ mission planners for programming the
ToO observations described in the paper. 
The present work is partly based on observations with INTEGRAL, and ESA project with instruments 
and science data centre funded by ESA member states (especially the PI countries: 
Denmark, France, Germany, Italy, Switzerland, Spain), and Poland, and with the participation of Russia and the USA, and on observations with \rxte. The NRAO is
a facility of the National Science Foundation operated under cooperative
agreement by Associated Universities, Inc. 
M.C.B. acknowledges support from the Faculty of the European Space Astronomy 
Center (ESAC). JR acknowledges partial funding from the European Community's Seventh
Framework Programme (FP7/2007-2013) under grant agreement number 
ITN 215212 "Black Hole Universe". JAT acknowledges partial support from NASA under Swift Guest Observer
grant NNX08AW35G. The FT North and South are maintained and 
operated by Las Cumbres Observatory Global Telescope Network. This research has made use of the
NASA Astrophysics Data System Abstract Service and of the SIMBAD database,
operated at the CDS, Strasbourg, France.

\end{acknowledgements}

\bibliographystyle{aa}
\bibliography{17684-gx_bib}

\begin{thebibliography}{90}
\expandafter\ifx\csname natexlab\endcsname\relax\def\natexlab#1{#1}\fi

\bibitem[{Arnaud(1996)}]{Arnaud:1996}
Arnaud, K.~A. 1996, ASP Conferences, 101, 17

\bibitem[{Belloni {et~al.}(2005)Belloni, Homan, Casella, van~der Klis, Nespoli,
  Lewin, Miller, \& M{\'e}ndez}]{Belloni:2005}
Belloni, T., Homan, J., Casella, P., {et~al.} 2005, A{\&}A, 440, 207

\bibitem[{Belloni {et~al.}(2001)Belloni, M{\'e}ndez, \&
  S{\'a}nchez-Fern{\'a}ndez}]{Belloni:2001}
Belloni, T., M{\'e}ndez, M., \& S{\'a}nchez-Fern{\'a}ndez, C. 2001, A{\&}A,
  372, 551

\bibitem[{{Belloni}(2010)}]{Bell10}
{Belloni}, T.~M. 2010, in Lecture Notes in Physics, Berlin Springer Verlag, ed.
  {T.~Belloni}, Vol. 794, 53

\bibitem[{{Bertin} \& {Arnouts}(1996)}]{Bertin96}
{Bertin}, E. \& {Arnouts}, S. 1996, \aaps, 117, 393

\bibitem[{{Buxton} {et~al.}(2010){Buxton}, {Dincer}, {Kalemci}, \&
  {Tomsick}}]{Bux10}
{Buxton}, M., {Dincer}, T., {Kalemci}, E., \& {Tomsick}, J. 2010, The
  Astronomer's Telegram, 2549, 1

\bibitem[{{Buxton} \& {Bailyn}(2004)}]{bux04}
{Buxton}, M.~M. \& {Bailyn}, C.~D. 2004, \apj, 615, 880

\bibitem[{{Cabanac} {et~al.}(2009){Cabanac}, {Fender}, {Dunn}, \&
  {K{\"o}rding}}]{cabanac}
{Cabanac}, C., {Fender}, R.~P., {Dunn}, R.~J.~H., \& {K{\"o}rding}, E.~G. 2009,
  \mnras, 396, 1415

\bibitem[{Cadolle~Bel {et~al.}(2008)Cadolle~Bel, Kuulkers, Barrag\'an,
  {et~al.}}]{CadolleBel:2008}
Cadolle~Bel, M., Kuulkers, E., Barrag\'an, L., {et~al.} 2008, Proceedings of "A
  Population Explosion", St Petersburg/FL, USA, 28 Oct-02 Nov 2007

\bibitem[{{Cadolle Bel} {et~al.}(2010){Cadolle Bel}, {Kuulkers}, {Ibarra},
  {Diaz Trigo}, {Tomsick}, {Rodriguez}, {Prat}, {Corbel}, {Russell},
  {Altamirano}, {Lewis}, {Bozzo}, {Turler}, \& {Ferrigno}}]{cad10}
{Cadolle Bel}, M., {Kuulkers}, E., {Ibarra}, A., {et~al.} 2010, The
  Astronomer's Telegram, 2573, 1

\bibitem[{{Cadolle Bel} {et~al.}(2009){Cadolle Bel}, {Prat}, {Rodriguez},
  {Rib{\'o}}, {Barrag{\'a}n}, {D'Avanzo}, {Hannikainen}, {Kuulkers}, {Campana},
  {Mold{\'o}n}, {Chaty}, {Zurita-Heras}, {Goldwurm}, \& {Goldoni}}]{cad09}
{Cadolle Bel}, M., {Prat}, L., {Rodriguez}, J., {et~al.} 2009, \aap, 501, 1

\bibitem[{Cadolle~Bel {et~al.}(2004)Cadolle~Bel, Rodriguez, Sizun, Farinelli,
  Santo, Goldwurm, Goldoni, Corbel, Parmar, Kuulkers, Ubertini, Capitanio,
  Roques, Frontera, Amati, \& Westergaard}]{CadolleBel:2004}
Cadolle~Bel, M., Rodriguez, J., Sizun, P., {et~al.} 2004, \aap, 426, 659

\bibitem[{{Cadolle Bel} {et~al.}(2006){Cadolle Bel}, {Sizun}, {Goldwurm},
  {Rodriguez}, {Laurent}, {Zdziarski}, {Foschini}, {Goldoni}, {Gouiff{\`e}s},
  {Malzac}, {Jourdain}, \& {Roques}}]{cad06}
{Cadolle Bel}, M., {Sizun}, P., {Goldwurm}, A., {et~al.} 2006, \aap, 446, 591

\bibitem[{{Casella} {et~al.}(2010){Casella}, {Maccarone}, {O'Brien}, {Fender},
  {Russell}, {van der Klis}, {Pe'Er}, {Maitra}, {Altamirano}, {Belloni},
  {Kanbach}, {Klein-Wolt}, {Mason}, {Soleri}, {Stefanescu}, {Wiersema}, \&
  {Wijnands}}]{casella10}
{Casella}, P., {Maccarone}, T.~J., {O'Brien}, K., {et~al.} 2010, \mnras, 404,
  L21

\bibitem[{Chaty {et~al.}(2003)Chaty, Haswell, Malzac, Hynes, Shrader, \&
  Cui}]{Chaty:2003}
Chaty, S., Haswell, C.~A., Malzac, J., {et~al.} 2003, \mnras, 346, 689

\bibitem[{Chen {et~al.}(1997)Chen, Shrader, \& Livio}]{Chen:1997}
Chen, W., Shrader, C.~R., \& Livio, M. 1997, \apj, 491, 312

\bibitem[{{Chincarini} {et~al.}(2003){Chincarini}, {Zerbi}, {Antonelli},
  {Conconi}, {Cutispoto}, {Covino}, {D'Alessio}, {de Ugarte Postigo},
  {Molinari}, {Nicastro}, {Tosti}, {Vitali}, {Mazzoleni}, {Sciuto}, {Stefanon},
  {Jordan}, {Burderi}, {Campana}, {Danziger}, {di Paola}, {Fernandez-Soto},
  {Fiore}, {Ghisellini}, {Goldoni}, {Israel}, {Lorenzetti}, {Mc Breen},
  {Masetti}, {Messina}, {Meurs}, {Monfardini}, {Palazzi}, {Paul}, {Pian},
  {Rodono}, {Stella}, {Tagliaferri}, {Testa}, \& {Vergani}}]{chin}
{Chincarini}, G., {Zerbi}, F., {Antonelli}, A., {et~al.} 2003, The Messenger,
  113, 40

\bibitem[{{Corbel} {et~al.}(2010{\natexlab{a}}){Corbel}, {Broderick},
  {Brocksopp}, {Tzioumis}, \& {.}}]{corb10a}
{Corbel}, S., {Broderick}, J., {Brocksopp}, C., {Tzioumis}, T., \& {.}, R.~F.
  2010{\natexlab{a}}, The Astronomer's Telegram, 2525, 1

\bibitem[{{Corbel} {et~al.}(2010{\natexlab{b}}){Corbel}, {Broderick},
  {Calvelo}, {Kaaret}, {Brocksopp}, {Tomsick}, {Orosz}, {Coriat}, {Fender}, \&
  {Tzioumis}}]{corb10b}
{Corbel}, S., {Broderick}, J., {Calvelo}, D., {et~al.} 2010{\natexlab{b}}, The
  Astronomer's Telegram, 2745, 1

\bibitem[{{Corbel} \& {Fender}(2002)}]{corbfen02}
{Corbel}, S. \& {Fender}, R.~P. 2002, \apjl, 573, L35

\bibitem[{Corbel {et~al.}(2000)Corbel, Fender, Tzioumis, Nowak, McIntyre,
  Durouchoux, \& Sood}]{Corbel:2000}
Corbel, S., Fender, R.~P., Tzioumis, A.~K., {et~al.} 2000, \aap, 359, 251

\bibitem[{Corbel {et~al.}(2003)Corbel, Nowak, Fender, Tzioumis, \&
  Markoff}]{Corbel:2003}
Corbel, S., Nowak, M.~A., Fender, R.~P., Tzioumis, A.~K., \& Markoff, S. 2003,
  \aap, 400, 1007

\bibitem[{{Coriat} {et~al.}(2009){Coriat}, {Corbel}, {Buxton}, {Bailyn},
  {Tomsick}, {K{\"o}rding}, \& {Kalemci}}]{coriat09}
{Coriat}, M., {Corbel}, S., {Buxton}, M.~M., {et~al.} 2009, \mnras, 400, 123

\bibitem[{{Covino} {et~al.}(2004){Covino}, {Stefanon}, {Sciuto},
  {Fernandez-Soto}, {Tosti}, {Zerbi}, {Chincarini}, {Antonelli}, {Conconi},
  {Cutispoto}, {Molinari}, {Nicastro}, \& {Rodono}}]{Cov}
{Covino}, S., {Stefanon}, M., {Sciuto}, G., {et~al.} 2004, in Presented at the
  Society of Photo-Optical Instrumentation Engineers (SPIE) Conference, Vol.
  5492, Society of Photo-Optical Instrumentation Engineers (SPIE) Conference
  Series, ed. {A.~F.~M.~Moorwood \& M.~Iye}, 1613--1622

\bibitem[{Done {et~al.}(2007)Done, Gierli{\'n}ski, \& Kubota}]{Done:2007}
Done, C., Gierli{\'n}ski, M., \& Kubota, A. 2007, \aapr, 15, 1

\bibitem[{Fender {et~al.}(2004)Fender, Belloni, \& Gallo}]{Fender:2004}
Fender, R.~P., Belloni, T.~M., \& Gallo, E. 2004, \mnras, 355, 1105

\bibitem[{{Fender} {et~al.}(2009){Fender}, {Homan}, \& {Belloni}}]{fender09}
{Fender}, R.~P., {Homan}, J., \& {Belloni}, T.~M. 2009, \mnras, 396, 1370

\bibitem[{Fitzpatrick(1999)}]{Fi99}
Fitzpatrick, E.~L. 1999, PASP, 111, 63

\bibitem[{{Frank} {et~al.}(2002){Frank}, {King}, \& {Raine}}]{Frank2002}
{Frank}, J., {King}, A., \& {Raine}, D.~J. 2002, {Accretion Power in
  Astrophysics: Third Edition}, ed. {Frank, J., King, A., \& Raine, D.~J.,
  Cambridge Univ. Press}, Vol.~21

\bibitem[{{Gallo} {et~al.}(2004){Gallo}, {Corbel}, {Fender}, {Maccarone}, \&
  {Tzioumis}}]{Gallo04}
{Gallo}, E., {Corbel}, S., {Fender}, R.~P., {Maccarone}, T.~J., \& {Tzioumis},
  A.~K. 2004, \mnras, 347, L52

\bibitem[{Gallo {et~al.}(2006)Gallo, Fender, Miller-Jones, Merloni, Jonker,
  Heinz, Maccarone, \& van~der Klis}]{Gallo:2006}
Gallo, E., Fender, R.~P., Miller-Jones, J. C.~A., {et~al.} 2006, \mnras, 370,
  1351

\bibitem[{Gallo {et~al.}(2003)Gallo, Fender, \& Pooley}]{Gallo:2003}
Gallo, E., Fender, R.~P., \& Pooley, G.~G. 2003, \mnras, 344, 60

\bibitem[{{Gandhi} {et~al.}(2010){Gandhi}, {Dhillon}, {Durant}, {Fabian},
  {Kubota}, {Makishima}, {Malzac}, {Marsh}, {Miller}, {Shahbaz}, {Spruit}, \&
  {Casella}}]{gand10}
{Gandhi}, P., {Dhillon}, V.~S., {Durant}, M., {et~al.} 2010, \mnras, 407, 2166

\bibitem[{Gierli{\'n}ski {et~al.}(2008)Gierli{\'n}ski, Done, \&
  Page}]{Gierlinski:2008}
Gierli{\'n}ski, M., Done, C., \& Page, K. 2008, \mnras, 388, 753

\bibitem[{{Gierli{\'n}ski} {et~al.}(2009){Gierli{\'n}ski}, {Done}, \&
  {Page}}]{gier2009}
{Gierli{\'n}ski}, M., {Done}, C., \& {Page}, K. 2009, \mnras, 392, 1106

\bibitem[{{Grove} {et~al.}(1998){Grove}, {Johnson}, {Kroeger}, {McNaron-Brown},
  {Skibo}, \& {Phlips}}]{Grove98}
{Grove}, J.~E., {Johnson}, W.~N., {Kroeger}, R.~A., {et~al.} 1998, \apj, 500,
  899

\bibitem[{Homan \& Belloni(2005)}]{Homan:2005}
Homan, J. \& Belloni, T. 2005, Astrophysics and Space Science, 300, 107

\bibitem[{{Homan} {et~al.}(2005){Homan}, {Buxton}, {Markoff}, {Bailyn},
  {Nespoli}, \& {Belloni}}]{Homan:2005b}
{Homan}, J., {Buxton}, M., {Markoff}, S., {et~al.} 2005, \apj, 624, 295

\bibitem[{Hynes(2005)}]{Hynes:2005}
Hynes, R.~I. 2005, \apj, 623, 1026

\bibitem[{{Hynes} {et~al.}(2003){Hynes}, {Steeghs}, {Casares}, {Charles}, \&
  {O'Brien}}]{hynes2003}
{Hynes}, R.~I., {Steeghs}, D., {Casares}, J., {Charles}, P.~A., \& {O'Brien},
  K. 2003, \apjl, 583, L95

\bibitem[{{Hynes} {et~al.}(2004){Hynes}, {Steeghs}, {Casares}, {Charles}, \&
  {O'Brien}}]{Hynes04}
{Hynes}, R.~I., {Steeghs}, D., {Casares}, J., {Charles}, P.~A., \& {O'Brien},
  K. 2004, \apj, 609, 317

\bibitem[{Jahoda {et~al.}(2006)Jahoda, Markwardt, Radeva, Rots, Stark, Swank,
  Strohmayer, \& Zhang}]{Jahoda:2006}
Jahoda, K., Markwardt, C.~B., Radeva, Y., {et~al.} 2006, \apjs, 163, 401

\bibitem[{{Jordi} {et~al.}(2006){Jordi}, {Grebel}, \& {Ammon}}]{jordet06}
{Jordi}, K., {Grebel}, E.~K., \& {Ammon}, K. 2006, \aap, 460, 339

\bibitem[{{Kalberla} {et~al.}(2005){Kalberla}, {Burton}, {Hartmann}, {Arnal},
  {Bajaja}, {Morras}, \& {P{\"o}ppel}}]{kalberla05}
{Kalberla}, P.~M.~W., {Burton}, W.~B., {Hartmann}, D., {et~al.} 2005, \aap,
  440, 775

\bibitem[{{Kataoka} {et~al.}(2008){Kataoka}, {Madejski}, {Sikora}, {Roming},
  {Chester}, {Grupe}, {Tsubuku}, {Sato}, {Kawai}, {Tosti}, {Impiombato},
  {Kovalev}, {Kovalev}, {Edwards}, {Wagner}, {Moderski}, {Stawarz},
  {Takahashi}, \& {Watanabe}}]{Kata08}
{Kataoka}, J., {Madejski}, G., {Sikora}, M., {et~al.} 2008, \apj, 672, 787

\bibitem[{{Kong}(2008)}]{Kong08}
{Kong}, A.~K.~H. 2008, The Astronomer's Telegram, 1588, 1

\bibitem[{Kuulkers {et~al.}(2007)Kuulkers, Shaw, Paizis, Chenevez, Brandt,
  Courvoisier, Domingo, Ebisawa, Kretschmar, Markwardt, Mowlavi, Oosterbroek,
  Orr, R{\'\i}squez, Sanchez-Fernandez, \& Wijnands}]{Kuulkers:2007}
Kuulkers, E., Shaw, S.~E., Paizis, A., {et~al.} 2007, \aap, 466, 595

\bibitem[{{Landolt}(1992)}]{land92}
{Landolt}, A.~U. 1992, \aj, 104, 372

\bibitem[{{Laurent} {et~al.}(2011){Laurent}, {Rodriguez}, {Wilms}, {Cadolle
  Bel}, {Pottschmidt}, \& {Grinberg}}]{laurent2011}
{Laurent}, P., {Rodriguez}, J., {Wilms}, J., {et~al.} 2011, Science, 332, 438

\bibitem[{{Laurent} \& {Titarchuk}(2007)}]{laurent07}
{Laurent}, P. \& {Titarchuk}, L. 2007, \apj, 656, 1056

\bibitem[{{Lewis} \& {Russell}(2009)}]{lew09}
{Lewis}, F. \& {Russell}, D.~M. 2009, The Astronomer's Telegram, 2270, 1

\bibitem[{{Lewis} {et~al.}(2010){Lewis}, {Russell}, \& {Cadolle Bel}}]{Lew10}
{Lewis}, F., {Russell}, D.~M., \& {Cadolle Bel}, M. 2010, The Astronomer's
  Telegram, 2459, 1

\bibitem[{{Lewis} {et~al.}(2008){Lewis}, {Russell}, {Fender}, {Roche}, \&
  {Clark}}]{lewiet08}
{Lewis}, F., {Russell}, D.~M., {Fender}, R.~P., {Roche}, P., \& {Clark}, J.~S.
  2008, ArXiv e-prints

\bibitem[{Markoff {et~al.}(2005)Markoff, Nowak, \& Wilms}]{Markoff:2005}
Markoff, S., Nowak, M.~A., \& Wilms, J. 2005, \apj, 635, 1203

\bibitem[{{Markwardt}(2009)}]{Mark09}
{Markwardt}, C.~B.~M.~N.~S.~J.~H.~S.~D.~A.~K.~Y.~Y. 2009, The Astronomer's
  Telegram, 1945, 1

\bibitem[{McClintock \& Remillard(2006)}]{McClintock:2006}
McClintock, J.~E. \& Remillard, R.~A. 2006, Black hole binaries, Compact
  stellar X-ray sources. Edited by Walter Lewin \& Michiel van der Klis:
  Cambridge University Press, 157

\bibitem[{Miller {et~al.}(2006)Miller, Homan, Steeghs, Rupen, Hunstead,
  Wijnands, Charles, \& Fabian}]{Miller:2006}
Miller, J.~M., Homan, J., Steeghs, D., {et~al.} 2006, \apj, 653, 525

\bibitem[{{Mitsuda} {et~al.}(1984){Mitsuda}, {Inoue}, {Koyama}, {Makishima},
  {Matsuoka}, {Ogawara}, {Suzuki}, {Tanaka}, {Shibazaki}, \&
  {Hirano}}]{Mitsuda:1984}
{Mitsuda}, K., {Inoue}, H., {Koyama}, K., {et~al.} 1984, \pasj, 36, 741

\bibitem[{{Motta} {et~al.}(2009){Motta}, {Belloni}, \& {Homan}}]{Motta09}
{Motta}, S., {Belloni}, T., \& {Homan}, J. 2009, \mnras, 400, 1603

\bibitem[{{Motta} {et~al.}(2010{\natexlab{a}}){Motta}, {Belloni}, \& {Mu{\~n}oz
  Darias}}]{motta10}
{Motta}, S., {Belloni}, T., \& {Mu{\~n}oz Darias}, T. 2010{\natexlab{a}}, The
  Astronomer's Telegram, 2545, 1

\bibitem[{{Motta} {et~al.}(2010{\natexlab{b}}){Motta}, {Belloni},
  {Mu{\~n}oz-Darias}, \& {Homan}}]{Motta10b}
{Motta}, S., {Belloni}, T., {Mu{\~n}oz-Darias}, T., \& {Homan}, J.
  2010{\natexlab{b}}, The Astronomer's Telegram, 2593, 1

\bibitem[{{Mu{\~n}oz-Darias} {et~al.}(2008){Mu{\~n}oz-Darias}, {Casares}, \&
  {Mart{\'{\i}}nez-Pais}}]{munoz2008}
{Mu{\~n}oz-Darias}, T., {Casares}, J., \& {Mart{\'{\i}}nez-Pais}, I.~G. 2008,
  \mnras, 385, 2205

\bibitem[{{Poole} {et~al.}(2008){Poole}, {Breeveld}, {Page}, {Landsman},
  {Holland}, {Roming}, {Kuin}, {Brown}, {Gronwall}, {Hunsberger}, {Koch},
  {Mason}, {Schady}, {vanden Berk}, {Blustin}, {Boyd}, {Broos}, {Carter},
  {Chester}, {Cucchiara}, {Hancock}, {Huckle}, {Immler}, {Ivanushkina},
  {Kennedy}, {Marshall}, {Morgan}, {Pandey}, {de Pasquale}, {Smith}, \&
  {Still}}]{poolet08}
{Poole}, T.~S., {Breeveld}, A.~A., {Page}, M.~J., {et~al.} 2008, \mnras, 383,
  627

\bibitem[{{Prat} {et~al.}(2010){Prat}, {Cadolle Bel}, {Terrier}, {Pavan},
  {Vovk}, {Rodriguez}, {Corbel}, {Coriat}, \& {Kuulkers}}]{Prat10}
{Prat}, L., {Cadolle Bel}, M., {Terrier}, R., {et~al.} 2010, The Astronomer's
  Telegram, 2455, 1

\bibitem[{{Rahoui} {et~al.}(2010){Rahoui}, {Chaty}, {Rodriguez}, {Fuchs},
  {Mirabel}, \& {Pooley}}]{Rahoui10}
{Rahoui}, F., {Chaty}, S., {Rodriguez}, J., {et~al.} 2010, \apj, 715, 1191

\bibitem[{Rodriguez {et~al.}(2008)Rodriguez, Shaw, Hannikainen, Belloni,
  Corbel, Bel, Chenevez, Prat, Kretschmar, Lehto, Mirabel, Paizis, Pooley,
  Tagger, Varni{\`e}re, Cabanac, \& Vilhu}]{Rodriguez:2008b}
Rodriguez, J., Shaw, S.~E., Hannikainen, D.~C., {et~al.} 2008, \apj, 675, 1449

\bibitem[{{Roming} {et~al.}(2005){Roming}, {Kennedy}, {Mason}, {Nousek}, {Ahr},
  {Bingham}, {Broos}, {Carter}, {Hancock}, {Huckle}, {Hunsberger}, {Kawakami},
  {Killough}, {Koch}, {McLelland}, {Smith}, {Smith}, {Soto}, {Boyd},
  {Breeveld}, {Holland}, {Ivanushkina}, {Pryzby}, {Still}, \&
  {Stock}}]{romiet05}
{Roming}, P.~W.~A., {Kennedy}, T.~E., {Mason}, K.~O., {et~al.} 2005, \ssr, 120,
  95

\bibitem[{{Russell} {et~al.}(2010){Russell}, {Buxton}, {Lewis}, \&
  {Altamirano}}]{Ru10}
{Russell}, D.~M., {Buxton}, M., {Lewis}, F., \& {Altamirano}, D. 2010, The
  Astronomer's Telegram, 2547, 1

\bibitem[{Russell {et~al.}(2006)Russell, Fender, Hynes, Brocksopp, Homan,
  Jonker, \& Buxton}]{Russell:2006}
Russell, D.~M., Fender, R.~P., Hynes, R.~I., {et~al.} 2006, \mnras, 371, 1334

\bibitem[{{Russell} {et~al.}(2007){Russell}, {Fender}, \& {Jonker}}]{russ07}
{Russell}, D.~M., {Fender}, R.~P., \& {Jonker}, P.~G. 2007, \mnras, 379, 1108

\bibitem[{{Russell} \& {Lewis}(2011{\natexlab{a}})}]{Russ11c}
{Russell}, D.~M. \& {Lewis}, F. 2011{\natexlab{a}}, The Astronomer's Telegram,
  3191, 1

\bibitem[{{Russell} \& {Lewis}(2011{\natexlab{b}})}]{Russ11d}
{Russell}, D.~M. \& {Lewis}, F. 2011{\natexlab{b}}, The Astronomer's Telegram,
  3383, 1

\bibitem[{{Russell} {et~al.}(2011){Russell}, {Maitra}, {Dunn}, \&
  {Fender}}]{Russ2011}
{Russell}, D.~M., {Maitra}, D., {Dunn}, R.~J.~H., \& {Fender}, R.~P. 2011,
  ArXiv e-prints

\bibitem[{Rykoff {et~al.}(2007)Rykoff, Miller, Steeghs, \&
  Torres}]{Rykoff:2007}
Rykoff, E.~S., Miller, J.~M., Steeghs, D., \& Torres, M. A.~P. 2007, \apj, 666,
  1129

\bibitem[{{Shahbaz} {et~al.}(2001){Shahbaz}, {Fender}, \& {Charles}}]{Shab01}
{Shahbaz}, T., {Fender}, R., \& {Charles}, P.~A. 2001, \aap, 376, L17

\bibitem[{{Shaposhnikov} \& {Tomsick}(2010)}]{shap10}
{Shaposhnikov}, N. \& {Tomsick}, J.~A. 2010, The Astronomer's Telegram, 2523, 1

\bibitem[{Tanaka \& Lewin(1995)}]{Tanaka:1995}
Tanaka, Y. \& Lewin, W. H.~G. 1995, in "X-ray Binaries", ed. Lewin, van
  Paradijs, \& van~den Heuvel (Cambridge University Press), 126

\bibitem[{Tanaka \& Shibazaki(1996)}]{Tanaka:1996}
Tanaka, Y. \& Shibazaki, N. 1996, \araa, 34, 607

\bibitem[{Titarchuk(1994)}]{Titarchuk:1994}
Titarchuk, L. 1994, \apj, 434, 570

\bibitem[{{Tomsick}(2010)}]{tomsick2010}
{Tomsick}, J.~A. 2010, The Astronomer's Telegram, 2384, 1

\bibitem[{{Tomsick} {et~al.}(2009){Tomsick}, {Yamaoka}, {Corbel}, {Kaaret},
  {Kalemci}, \& {Migliari}}]{tomsick09}
{Tomsick}, J.~A., {Yamaoka}, K., {Corbel}, S., {et~al.} 2009, \apjl, 707, L87

\bibitem[{{van Paradijs} \& {McClintock}(1994)}]{vanpar1994}
{van Paradijs}, J. \& {McClintock}, J.~E. 1994, \aap, 290, 133

\bibitem[{{Wilms} {et~al.}(2000){Wilms}, {Allen}, \& {McCray}}]{wilms00}
{Wilms}, J., {Allen}, A., \& {McCray}, R. 2000, \apj, 542, 914

\bibitem[{{Wu} {et~al.}(2010){Wu}, {Yu}, {Yan}, {Sun}, \& {Li}}]{Wu10}
{Wu}, Y.~X., {Yu}, W., {Yan}, Z., {Sun}, L., \& {Li}, T.~P. 2010, \aap, 512,
  A32+

\bibitem[{{Yamaoka} {et~al.}(2010){Yamaoka}, {Sugizaki}, {Nakahira}, {Mihara},
  {Kohama}, {Nakagawa}, {Yamamoto}, {Matsuoka}, {Kawasaki}, {Ueno}, {Tomida},
  {Suzuki}, {Ishikawa}, {Kawai}, {Morii}, {Sugimori}, {Yoshida}, {Tsunemi},
  {Kimura}, {Negoro}, {Nakajima}, {Ishiwata}, {Miyoshi}, {Ozawa}, {Ueda},
  {Isobe}, {Eguchi}, {Hiroi}, \& {Daikyuji}}]{yam}
{Yamaoka}, K., {Sugizaki}, M., {Nakahira}, S., {et~al.} 2010, The Astronomer's
  Telegram, 2380, 1

\bibitem[{{Yu}(2010)}]{yu10}
{Yu}, W. 2010, The Astronomer's Telegram, 2556, 1

\bibitem[{{Zdziarski} {et~al.}(2004){Zdziarski}, {Gierli{\'n}ski},
  {Miko{\l}ajewska}, {Wardzi{\'n}ski}, {Smith}, {Harmon}, \&
  {Kitamoto}}]{zdz04}
{Zdziarski}, A.~A., {Gierli{\'n}ski}, M., {Miko{\l}ajewska}, J., {et~al.} 2004,
  \mnras, 351, 791

\bibitem[{{Zdziarski} {et~al.}(1998){Zdziarski}, {Poutanen}, {Mikolajewska},
  {Gierlinski}, {Ebisawa}, \& {Johnson}}]{Zd98}
{Zdziarski}, A.~A., {Poutanen}, J., {Mikolajewska}, J., {et~al.} 1998, \mnras,
  301, 435

\bibitem[{{Zerbi} \& {The Rem Team}(2001)}]{Zerb}
{Zerbi}, F.~M. \& {The Rem Team}. 2001, in Astronomische Gesellschaft Meeting
  Abstracts, Vol.~18, Astronomische Gesellschaft Meeting Abstracts, ed.
  {E.~R.~Schielicke}, J101+

\bibitem[{Zurita {et~al.}(2006)Zurita, Torres, Steeghs, Rodr{\'\i}guez-Gil,
  Mu{\~n}oz-Darias, Casares, Shahbaz, Mart{\'\i}nez-Pais, Zhao, Garcia,
  Piccioni, Bartolini, Guarnieri, Bloom, Blake, Falco, Szentgyorgyi, \&
  Skrutskie}]{Zurita:2006}
Zurita, C., Torres, M. A.~P., Steeghs, D., {et~al.} 2006, \apj, 644, 432

\end{thebibliography}

\end{document}